\documentclass[useAMS,usenatbib,a4paper,fleqn]{mn2e}
\usepackage{mathptmx}
\usepackage{ifthen}
\usepackage{hyperref}
\usepackage{etoolbox}
\usepackage{xcolor}
\usepackage{graphicx} 
\usepackage{xspace}
\usepackage{amssymb}
\usepackage{amsmath}
\usepackage{paralist}

\renewcommand{\arcmin}{\ensuremath{\,{\rm arcmin}}\xspace}

\newcommand{\ie}{{\it i.e.}\xspace}
\newcommand{\eg}{{\it e.g.}\xspace}
\newcommand{\cf}{{\it cf.}\xspace}

\newcommand{\kpc}{\ensuremath{\,{\rm kpc}}\xspace}
\newcommand{\pc}{\ensuremath{\,{\rm pc}}\xspace}
\newcommand{\Gyr}{\ensuremath{\,{\rm Gyr}}\xspace}
\newcommand{\mags}{\ensuremath{\,{\rm mag}}\xspace}
\newcommand{\Kdered}{\ensuremath{K_{s0}}\xspace}
\newcommand{\K}{\ensuremath{{K_s}}\xspace}
\newcommand{\Kband}{\K-band\xspace}
\newcommand{\MK}{\ensuremath{{M_K}}\xspace}
\newcommand{\MKRC}{\ensuremath{{M_{K,{\rm RC}}}}\xspace}
\newcommand{\MKRGBB}{\ensuremath{{M_{K,{\rm RGBB}}}}\xspace}
\newcommand{\R}{\ensuremath{ {R_{0}} }\xspace}

\newcommand{\JK}{\ensuremath{J-\K}\xspace}
\newcommand{\Ak}{\ensuremath{A_{K_s}}\xspace}
\newcommand{\FeH}{\ensuremath{[{\rm Fe/H}]}\xspace}
\newcommand{\nbin}{{\ensuremath{N_{\rm bin}}\xspace}}
\newcommand{\rhorc}{{\ensuremath{\rho_{\rm RC}}\xspace}}
\newcommand{\Phirc}{{\ensuremath{\Phi_{\rm RC}}\xspace}}
\newcommand{\solid}{{\ensuremath{\Delta \Omega\,}\xspace}}

\newcommand{\eqn}[1]{{equation \ref{eq:#1}\xspace}}
\newcommand{\Eqn}[1]{{Equation \ref{eq:#1}\xspace}}
\newcommand{\fig}[1]{{figure \ref{fig:#1}\xspace}}
\newcommand{\Fig}[1]{{Figure \ref{fig:#1}\xspace}}
\newcommand{\sect}[1]{{section \ref{sec:#1}\xspace}}

\title[The 3D Density of the Galactic Bulge]{Mapping the three-dimensional density of the Galactic bulge with VVV red clump stars}
 
\author[Wegg \& Gerhard]
{Christopher Wegg$^{1}$\thanks{E-mail: wegg@mpe.mpg.de} and Ortwin Gerhard$^{1}$\\
$^1$Max-Planck-Institut f\"ur Extraterrestrische Physik, Giessenbachstrasse, 85748 Garching, Germany.}

\pagerange{\pageref{firstpage}--\pageref{lastpage}} \pubyear{2012}
 
\begin{document}
\label{firstpage}
\maketitle

\begin{abstract}
The inner Milky Way is dominated by a boxy, triaxial bulge which
is believed to have formed through disk instability processes.
Despite its proximity, its large-scale properties are still not very well known, 
due to our position in the obscuring Galactic disk.

Here we make a measurement of the three-dimensional density
distribution of the Galactic bulge using red clump giants
identified in DR1 of the VVV survey. Our density map covers the inner
$(2.2\times1.4\times1.1)\kpc$ of the bulge/bar. Line-of-sight density distributions
are estimated by deconvolving extinction and completeness corrected
\Kband magnitude distributions. In constructing our measurement, we assume
that the three-dimensional bulge is 8-fold mirror triaxially
symmetric.  In doing so we measure the angle of the bar-bulge to the
line-of-sight to be $(27\pm 2)\degr$, where the dominant error is
systematic arising from the details of the deconvolution process.

The resulting density distribution shows a highly elongated bar with projected
axis ratios $\approx (1:2.1)$ for isophotes reaching $\sim 2\kpc$ along the major
axis. Along the bar axes the density falls off roughly
exponentially, with axis ratios $(10:6.3:2.6)$ and exponential
scale-lengths $(0.70:0.44:0.18)\kpc$. From about 400 pc above the Galactic plane, the
bulge density distribution displays a prominent X-structure. Overall,
the density distribution of the Galactic bulge is characteristic for a
strongly boxy/peanut shaped bulge within a barred galaxy.
\end{abstract}

\section{Introduction}

Red clump (RC) giant stars have been used as standard candles to probe the Galactic bulge on numerous occasions, beginning with \citet{Stanek:94,Stanek:97}, who confirmed evidence from gas kinematics and NIR photometry \citep{Binney:91,Weiland:94} that the Galactic bulge hosts a triaxial structure. 

While many works have considered only the position of the peak of the RC in the luminosity function \citet{Stanek:97} and \citet{Rattenbury:07} fitted the full three dimensional density models of \citet{Dwek:95} to OGLE-II data.
More recently \citet{Nataf:10} using OGLE-III data, and \citet{McWilliam:10} using 2MASS, demonstrated that at high latitudes the RC splits into two as a result of the X-shaped structure characteristic for boxy/peanut bulges in barred galaxies (\eg \citealt{Inma:06} from N-body simulations and \citealt{Bureau:06} from observations).
\citet{Saito:11} explored the global structure using 2MASS RC stars, further confirming an X-shape in the bulge.

This paper expands on this earlier work by utilising DR1 of the VVV survey \citep{Saito:12}. The depth exceeds that of 2MASS by $\sim 4\mags$ and therefore allows the detection of the entire RC of the bulge in all but the most highly extincted regions.

In this work, rather than estimate the parameters of analytic densities, we infer a three dimensional density map of the bulge by inverting VVV star counts.
The methods in this work most closely resemble that of \citet{LopezCorredoira:00,LopezCorredoira:05} who derived three dimensional density distributions using a luminosity function of stars brighter than the RC in 2MASS data. The RC however is better for this purpose due to its small variation in absolute magnitude, and the newly available deeper VVV survey data allows the RC to be probed across the Galactic bulge for latitudes $\left| b \right| \gtrsim 1 \degr$ \citep{gonzalez:11a}.

This paper proceeds as follows: In section \ref{sec:magdists} we describe the process of constructing extinction and completeness corrected \K-band magnitude distributions for RC stars. In section \ref{sec:losrho} we describe how line-of-sight RC densities are extracted from these magnitude distributions by deconvolving the intrinsic RC luminosity function. In section \ref{sec:3drho} we use the line-of-sight densities to construct the three-dimensional density assuming 8-fold mirror symmetry. In section \ref{sec:dwek} we test that our code and methods are able to reconstruct a synthetic bulge. In section \ref{sec:caveats} we estimate systematic effects on our model by varying parameters from their fiducial values. In section \ref{sec:axis} we estimate axis ratios and projections of our density model. We discuss the properties of the density map and conclude in section \ref{sec:conc}.

\begin{figure*}
\includegraphics[width=\textwidth]{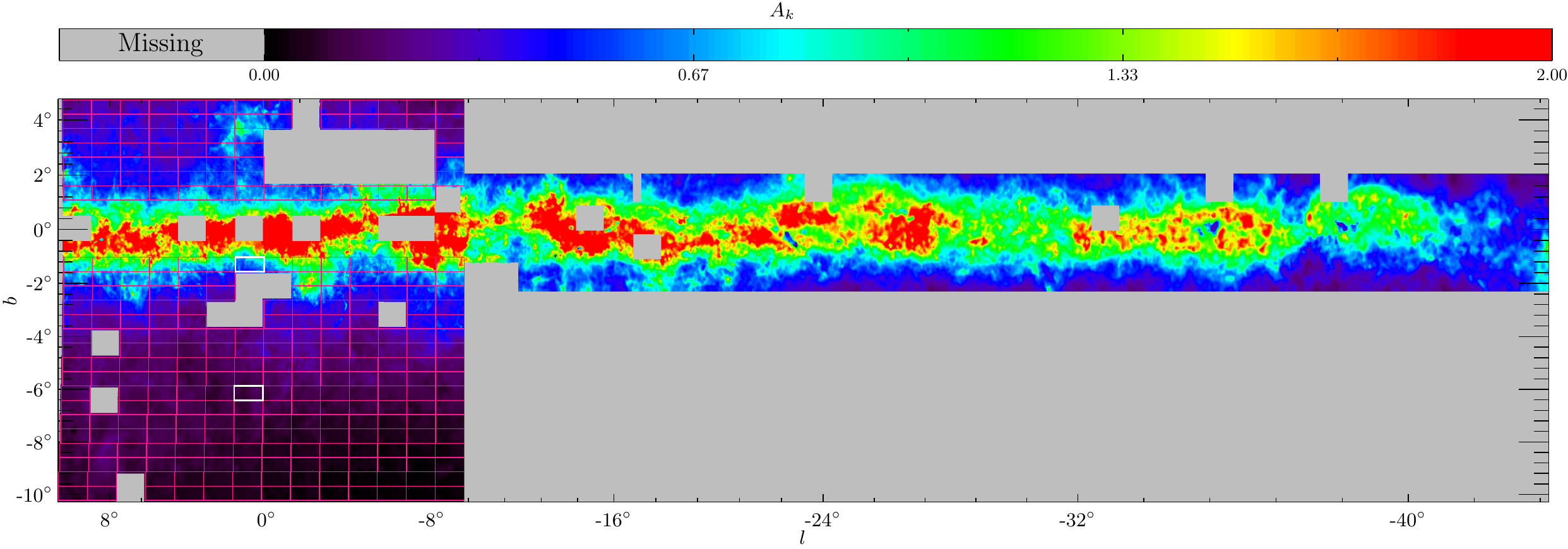}
\caption{The extinction map, $\Ak(l,b)$, constructed using VVV DR1 data using a similar method to \citet{gonzalez:11b,gonzalez:12}. The extinction map is used to compute the extinction corrected \K-band magnitude distributions. Grey regions are regions where the extinction could not be calculated because it was not surveyed in the $J$ and \K bands in VVV DR1. Pink outlines the fields analysed in this work. The fields plotted in \fig{catalogconstruct} are outlined in white.}
\label{fig:extmap}
\end{figure*}

\begin{figure}
\includegraphics[width=84mm]{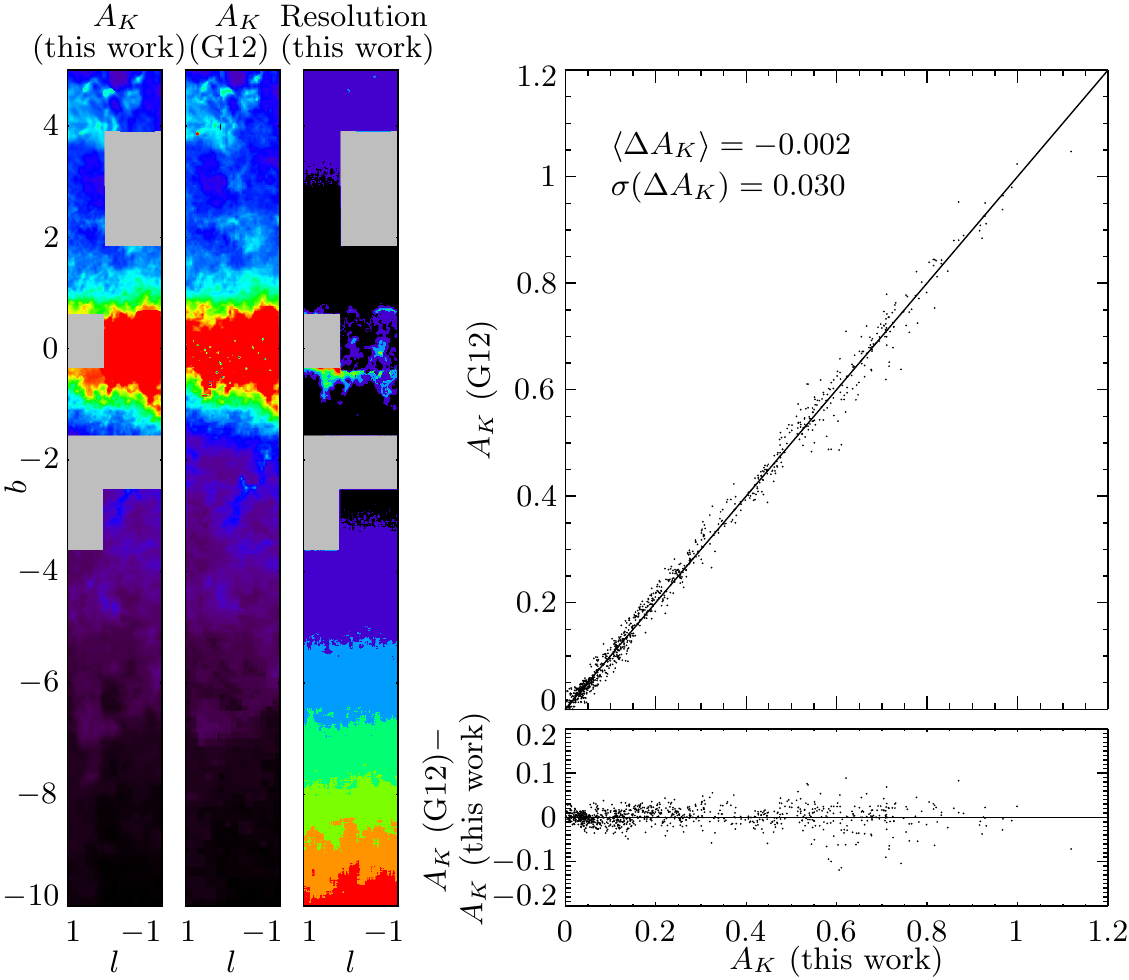}
\caption{Comparison of the extinction map produced in this work with that of \citet[][G12]{gonzalez:12}. On the left side we graphically compare a slice along the minor axis (the scale is equivalent to that in \fig{extmap}). On the right side we compare 1000 points randomly selected from this slice with $\left| b \right| > 1\degr$. The mean difference in extinction, considering all pixels across this region, is $-0.002\mags$, while the standard deviation is $0.030\mags$. We also show the adaptive resolution of our map for this slice, which is the primary difference between the extinction calculated here and in \citet{gonzalez:12}. The resolution, defined as the size of the box from which stars are selected in that pixel, ranges from 3\arcmin (black) to 15\arcmin (red) in steps of 2\arcmin.
\label{fig:akcomp}}
\end{figure}


\section{Magnitude Distributions}
\label{sec:magdists}

\subsection{Extinction Correction}

First an extinction corrected catalog of \Kband magnitudes is constructed, largely following the method  of \citet{gonzalez:11b}. We briefly summarise the method here.

Before beginning extinction map construction, the VVV DR1 zero-points are converted to zero-points based on 2MASS. The fiducial VVV DR1 zero points result in a field-to-field scatter of up to 0.1\mags in the position of the peak of the RC luminosity function at low galactic latitudes. Instead, in the same manner as \citet{gonzalez:11b}, the zero points are re-estimated for each field by cross matching bright but unsaturated VVV stars with 2MASS. This results in reduced field-to-field variation in the luminosity function in crowded low latitude regions.

The extinction corrected \Kband magnitudes for each star, $\Kdered = \K - \Ak(l,b)$, are calculated using extinction maps,  $\Ak(l,b)$, constructed from the data. We assume that the entirety of the extinction occurs between us and the bulge, and a negligible amount occurs within the bulge.

The extinction map is constructed by estimating the shift of the red clump in $\JK$ on a $1.0 \times 1.0 \arcmin$ grid. In each extinction map pixel the colour of the red clump, $\left< \JK \right>_{\rm RC}$, was estimated via a sigma-clipped mean. If less than 200 red clump stars are found in a pixel then it is combined with surrounding pixels until 200 red clump stars are included in the colour estimate for that pixel. In this manner the resolution of the extinction map is adaptive, ranging from $3\arcmin$ where the surface density of RC stars is highest within $\sim 3 \degr$ of the Galactic center, to $15\arcmin$ at $b\sim -10 \degr$.

The color of the red-clump, $\left< \JK \right>_{\rm RC}$, is converted to a reddening map of $E(\JK)$ using the position of the red clump in Baade's window (as in \citealt{gonzalez:11a}):
\begin{equation}
E(\JK) = \left< \JK \right>_{\rm RC} - \left< \JK \right>_{\rm RC,0} ~,
\end{equation}
where we adopt $\left< \JK \right>_{\rm RC,0} =0.674$ as the dereddened, intrinsic \JK colour of the red clump as measured by \citet{gonzalez:11a} in the direction of Baade's window.

\begin{figure*}
\includegraphics[width=84mm]{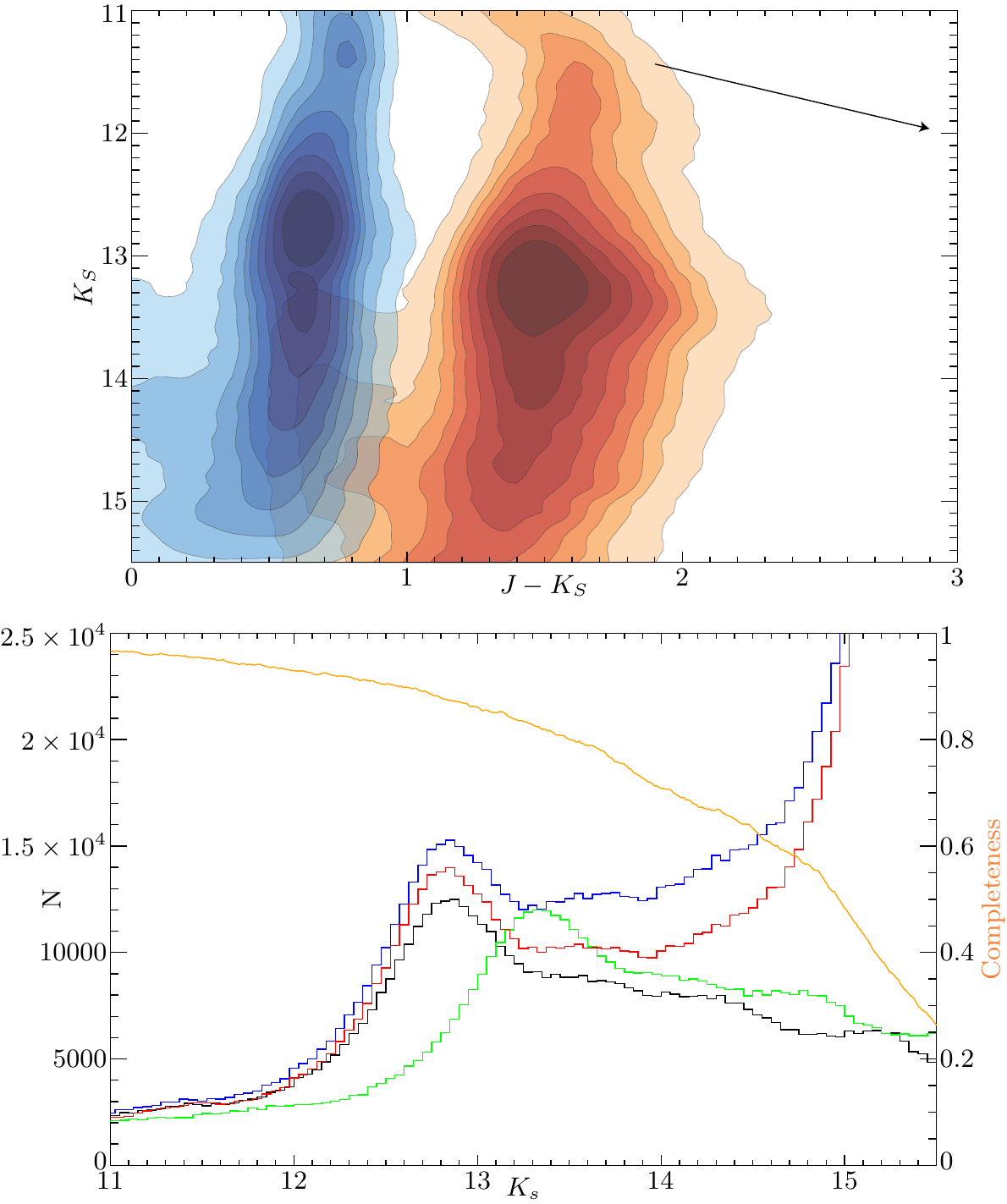}
\includegraphics[width=84mm]{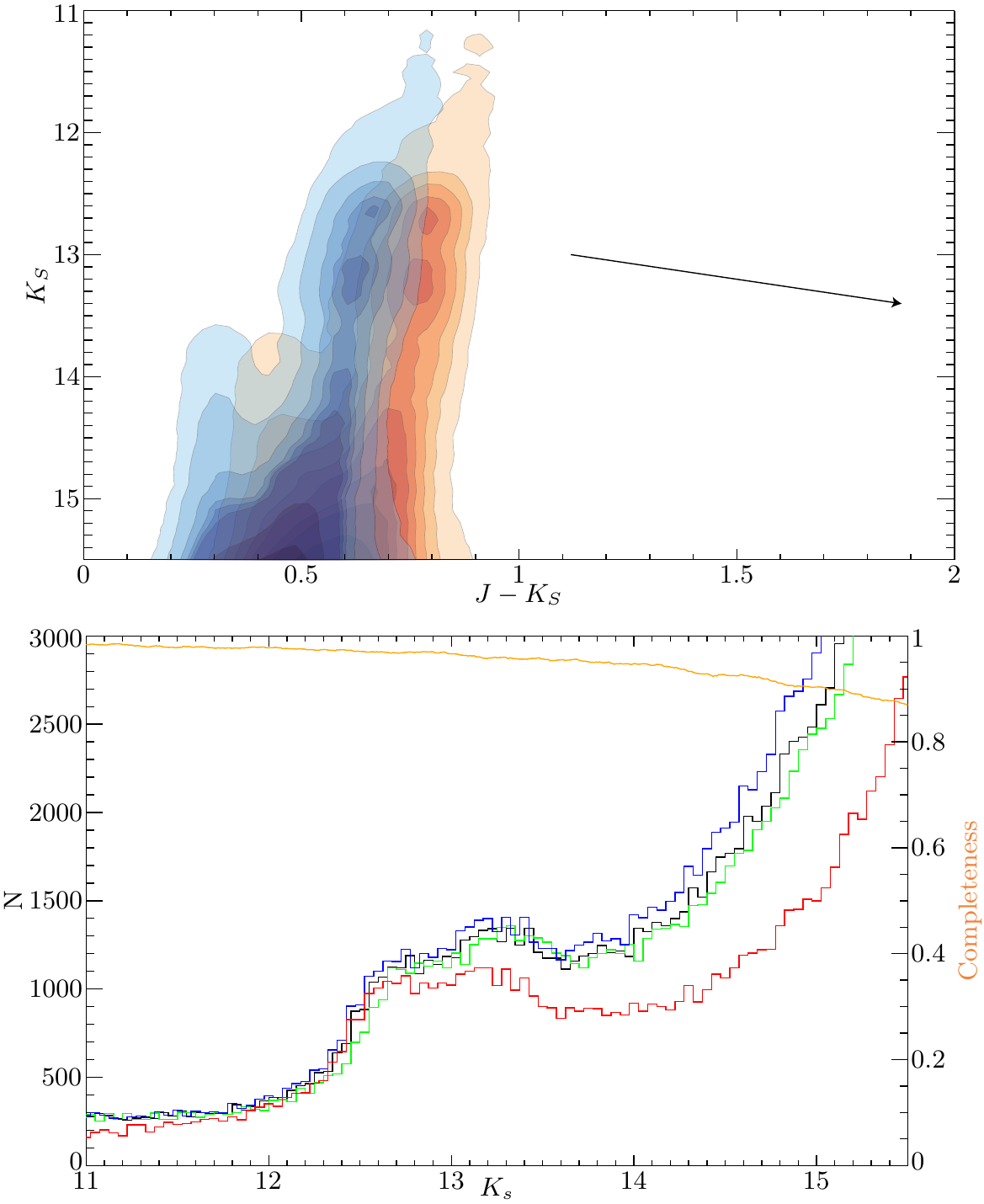}
\caption{Summary of catalog construction for two fields: The left column shows one of the most extincted fields considered lying at $(l,b)\approx (1,1)\degr$. The right column shows a field at $(l,b)\approx (-6,1)\degr$ which has only moderate extinction and displays the split red clump. The fields have an area $(\Delta l,\Delta b)\approx(1.5,0.5)\degr$ and are outlined in white in \fig{extmap}. In each column the top panel shows the colour magnitude diagram for the pre-extinction corrected stars in red, and the extinction corrected stars in blue. The iso-density contours contain $10,20,30....90\%$ of all stars in the field, and the extinction correction vector is the assumed extinction law \citep{Nishiyama:06}. The lower panel shows the \K-band magnitude distributions from this field in 0.05\mags bins. Green is the raw extincted magnitude distribution, black is extinction corrected, and blue is extinction and completeness corrected. The red curve is extinction and completeness corrected and includes a colour-cut: excluding stars with colours more than $3\sigma$ from the red clump. In addition, for the red curve, stars with extincted \K-band magnitude brighter than 12 were replaced with 2MASS data. The orange curve shows the calculated completeness, plotted on the right hand axis.
}
\label{fig:catalogconstruct}
\end{figure*}

Finally the extinction map $\Ak(l,b)$ is calculated from the map of $E(\JK)$ using the extinction law from \citet{Nishiyama:06}. This results in $\Ak = 0.528\,E(\JK)$ \citep{gonzalez:12}. We adopt the \citet{Nishiyama:06} extinction law since this was measured in the direction of the galactic bulge, and appears to more accurately reproduce the extinction of red clump stars in VVV \citep[see Fig 4 of][]{gonzalez:12}. In section \ref{sec:caveats} we investigate the consequences of adopting a different extinction law.

The resultant extinction map, $\Ak(l,b)$, is shown in \fig{extmap}. In \fig{akcomp} we show a comparison of the extinction map produced in this work with \citet{gonzalez:12}. The agreement is very good: over the region compared there is a negligible offset, and a standard deviation of only $0.03\mags$. The small differences could be explained by the adaptive resolution method used here, and differences in the details of the red clump colour fitting.

Adopting a constant red clump colour, $\left< \JK \right>_{\rm RC,0}$, is an approximation: The intrinsic \JK colour of the red clump is expected to be a function of metallicity \citep{Salaris:02,CabreraLavers:07}, and there is a gradient in the metallicity of the bulge \citep[\eg][]{Zoccali:08,Ness:13b,Gonzalez:13}. However the effect of this is negligible for the purposes of this work. The standard deviation of red clump stars in \JK in low extinction fields is $\approx 0.05$. These fields have a metallicity distribution with standard deviation $\sigma ({\FeH}) \approx 0.4 \mags$ \citep{Zoccali:08}. The range of mean metallicities over the area considered is $\Delta \FeH \approx 0.5$ \citep{Gonzalez:13} therefore, for the adopted extinction law, this corresponds to a variation in \Ak of only $0.528\times0.05\times0.5/0.4 \approx 0.03$.

\subsection{Completeness}

For fields close to the galactic plane, which are the most crowded and extincted fields, completeness becomes important at the magnitude of the red clump, despite the increased depth of VVV over previous surveys. To investigate and ultimately correct for completeness we performed artificial star tests using \textsc{daophot}. A Gaussian model for the PSF was constructed for each $\sim 20\arcmin\times20\arcmin$ section of each field. Using the PSF model 20,000 stars were added to each \K-band image at random positions with \K between 11 and 18\mags. The new image was then re-run though the CASU VISTA photometry pipeline software \textsc{imcore} using the same detection and photometry settings as VVV DR1. The artificial stars are considered detected if a source was detected within 1 pixel and $1\mags$ of the artificial star, although the completeness is insensitive to these choices. This entire process is repeated 5 times per image to reduce the statistical error in the completeness.

To correct for completeness then each star in the catalog is assigned a probability of detection given its extincted magnitude and the \K-band image in which it lies. This is calculated by estimating the probability of detection for the artificial stars within $\pm0.2\mags$. These probabilities are the completenesses shown in the lower panel of \fig{catalogconstruct}.  When constructing the magnitude distributions that follow, we do not use the raw number of stars in each magnitude bin, but instead sum the reciprocal of their computed completenesses.

The results of the extinction and completeness correction process are shown for two fields in \fig{catalogconstruct}.

\subsection{Catalog Construction}

To create a final catalog the individual catalogs from each VVV field are merged. In regions where VVV fields overlap the field closest to the center of the VISTA field of view is selected.

In order to reduce contamination from stars not part of the RC we make a colour-cut. We choose not to make a constant colour-cut in \JK, but instead reject more than $3\sigma_{\rm RC,JK}$ from the red clump, where $\sigma_{\rm RC,JK}$ is the standard deviation in \JK colour of the red clump as a function of $(l,b)$ measured during extinction map calculation. \Kband stars without a measured $J$-band companion are retained so as to not alter the completeness calculation. The result of this colour cut is shown as the red histogram in \fig{catalogconstruct}.

In addition regions within 3 half-light radii (or 6\arcmin when unavailable) of the center of any of the globular clusters listed in the catalog of \citet[][2010 edition]{Harris:96} were removed from further analysis, similarly to \citet{Nataf:13}.

Brighter than $\K\approx12$ the VVV the catalog begins to suffer from saturation and non-linearity \citep{Gonzalez:13}. When constructing luminosity functions from the catalog we therefore use 2MASS for $\K<12$. To ensure continuity in the luminosity function across this region we apply a constant fractional correction to the 2MASS luminosity function by requiring that the number of detections between 11.9 and 12.1 are equal. This correction is typically less than 5\%.


\section{Line of Sight Density Distribution}
\label{sec:losrho}

Before proceeding the extinction and completeness corrected catalog described in the previous section is divided into disjoint regions. These fields are aligned with the VVV images, but with each VVV image divided into two different latitude fields that are analysed separately. The resultant fields are $\approx 1.5^{\circ}\times0.5^{\circ}$ in size. We choose to make fields more closely spaced in latitude than longitude since the changes in magnitude distributions are more rapid in this direction as a result of the comparatively smaller scale height. At latitudes below $|b| < 1^{\circ}$ the extinction and completeness corrections at the magnitude of the red clump are too high for reliable analysis and we do not consider fields whose centres lie in this region. In addition we consider only VVV bulge fields (\ie $|l|\lesssim 10\degr$). The resultant fields are plotted in pink over the extinction map in \fig{extmap}.

Each field is considered to be a `pencil beam' with magnitude distribution, $N(\Kdered)$ \ie the number of stars between \Kdered and $\Kdered + d\K$ is $N(\Kdered) d\K$. This distribution arises from the equation of stellar statistics \citep[\eg][]{LopezCorredoira:00}:
\begin{equation}
N(\Kdered) = \solid \sum_i \int \Phi_i(\Kdered - 5 \log [ r/10\pc] ) \rho_i ( r )  r^2 \, dr ~ ,
\label{eq:stellarstats1}
\end{equation}
where $\Delta \Omega$ is the solid angle of the field, and the sum runs over the different populations, each with normalised luminosity function $\Phi_i(M_\K$), and line-of-sight density $\rho_i ( r )$. 

\begin{figure}
\includegraphics[width=84mm]{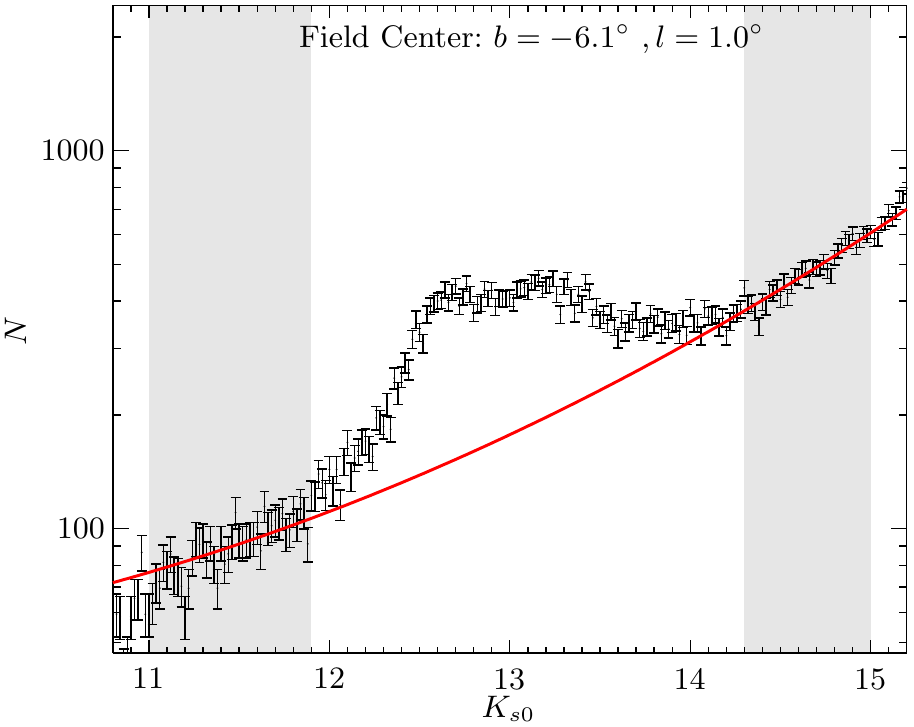}
\includegraphics[width=84mm]{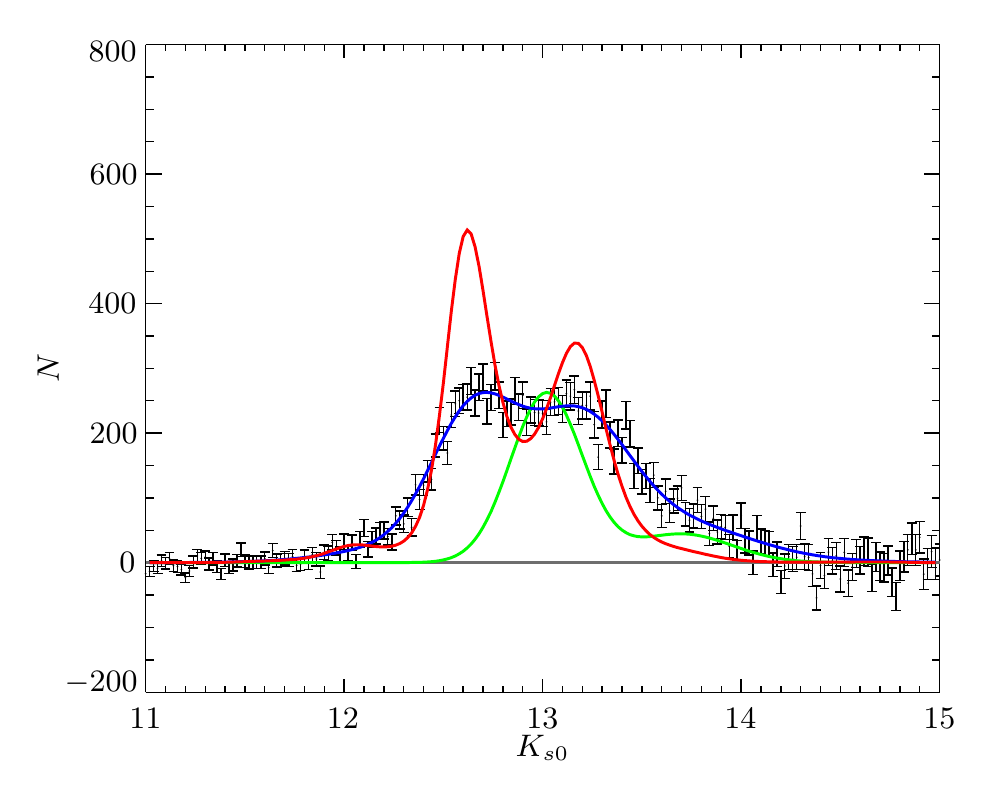}
\caption{Deconvolution process and extraction of the resultant line-of-sight density for one field. The sightline shown corresponds to the upper half of field b250, shown as the right plots of \fig{catalogconstruct}. In the upper figure we show the background fitting: A second order polynomial is fitted to the logarithm of the magnitude distribution in the shaded grey areas. This function is shown in red. In the lower panel we show the deconvolution: The density ($\Delta \equiv \solid (\ln 10/5) \rhorc r^3 $) in red, the assumed luminosity function ($\Phirc$) in green, and the resultant re-convolved model ($\Delta \star \Phirc$) in blue. In this case the iterative deconvolution stopped after 62 iterations because \eqn{stoppingcondition} was satisfied. The quality of the fit can be judged by comparing the re-convolved model (blue) with the data.  To allow easier comparison of $\Delta$ it was re-centered while plotting such that the red clump was effectively placed at $\MKRC=0$. Additionally while plotting $\Phirc$ was arbitrarily rescaled from the normalised form used in the analysis.}
\label{fig:deconvtest} 
\end{figure}

In fields away from the bulge the magnitude distribution is smooth on the scales of the bulge: a distance spread of 2\kpc at 8\kpc corresponds to 1.1\mags at  $\K = 13$ --- the magnitude of the red clump at $\sim 8\kpc$.
To proceed we therefore assume that populations other than red clump stars result in a smoothly varying `background' consisting primarily of first ascent giants and disk stars, and consider red clump stars separately:
\begin{equation}
N(\Kdered) = B(\Kdered) + \solid \int \rhorc(r) \Phirc(\Kdered - 5 \log [ r/10\pc] ) r^2 \, dr ~.
\label{eq:stellarstats}
\end{equation}
The exception to this assumption is red giant branch bump (RGBB), which is slight fainter than the red clump, with a similar dispersion in magnitude \citep{Nataf:11}. We therefore incorporate the RGBB as a secondary peak in the luminosity function, $\Phirc$, as described in section \ref{sec:lf}. We do not consider the asymptotic giant branch bump (AGBB) owing to its relatively small size \citep[only $\approx 3\%$ of the red clump,][]{Nataf:13}.

The fiducial form used for the background is a quadratic in $\log B$:
\begin{equation}
\log B(\Kdered) = a + b (\Kdered-13) + c (\Kdered-13)^2 ~.
\label{eq:bkgd}
\end{equation}
We investigate the consequences of changing this functional form in section \ref{sec:caveats}. We fit this background function over two regions either side of the RC. Specifically we use \K from  11 to 11.9 and 14.3 to 15\mags. This background fitting is shown for one field in the upper panel of \fig{deconvtest}. We find that the coefficient $c$ is generally small in comparison to $b$, but statistically significant (typically $\sim 0.02$ compared to $\approx 0.28$). 

There are two exception to this background fitting: fields with $l \geq 5.5\degr$,  and fields with $|b| \leq 2\degr$. The fields with $|b| \leq 2\degr$ have higher extinction and crowding, and as result uncertainties in completeness correction can result in spurious curvature to the background with these fitting choices. This can be seen as a steepening of the \K-band magnitude distribution near 15 in the field shown on the left in \fig{catalogconstruct}. For these fields we therefore set $c=0$ and restrict the background fit to $\K \leq 14.5$. Fields with $l \geq 5.5\degr$ have a brighter red clump distribution and we therefore restrict the background fitting at the bright end to the range 11--11.7\mags.

We plot background subtracted histograms for constant latitude slices in \fig{histslices}. These histograms are broadly comparable to figure 3 of \citet{Saito:11} using 2MASS data.

\subsection{Luminosity Function}
\label{sec:lf}

\begin{figure}
\includegraphics[width=84mm]{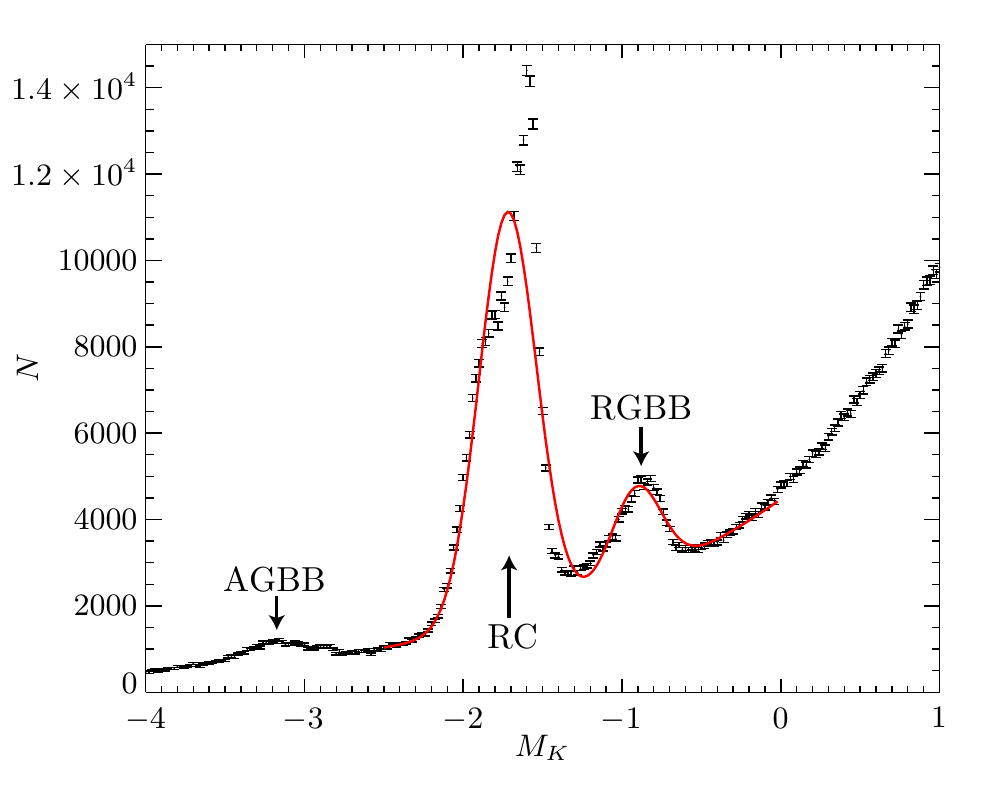}
\caption{Monte-Carlo simulation of the luminosity functions for the metallicity distribution in Baade's window measured by \citet{Zoccali:08} using the $\alpha$-enhanced \textsc{BaSTI} isochrones \citep{Pietrinferni:04} at 10\,Gyr and a Salpeter IMF. In red we plot our assumed luminosity function together with a fitted logarithmic background. The labeled features in the luminosity function are the red clump (RC), the red giant branch bump (RGBB) and the asymptotic giant branch bump (AGBB).}
\label{fig:fakelf}
\end{figure}

As described above, we consider the luminosity function to be deconvolved as consisting only of red clump (RC) stars and the red giant branch bump (RGBB), the remainder of the stars being a `background'. 
In the I-band, the RGBB is 0.74 mag fainter than the RC, and the
number of stars in the RGBB is $\sim 20\%$ of RC stars \citep{Nataf:13}.  
Thus the RGBB is important for deconvolving the bulge density,
particularly for sight-lines with a split red clump. Following \citet{Nataf:13}, we model the luminosity function of both the RC and the RGBB as
Gaussians.  Because the \Kband RGBB for the bulge has not been
well-studied, we perform a simple Monte Carlo simulation to estimate
the mean and standard deviation of both Gaussians.
We simulate stars drawn from the metallicity distribution function (MDF) in the direction of Baade's window measured by \citet{Zoccali:08}, with age $10\Gyr$, and draw masses from an initial mass function
(IMF). We use a Salpeter IMF, which is equivalent to a \citet{Kroupa:2001} IMF for the range of masses consider. From each metallicity and initial mass we then measure a theoretical \K-band absolute magnitude, \MK, by interpolating from the $\alpha$-enhanced \textsc{BaSTI} isochrones at 10\,Gyr \citep{Pietrinferni:04}. This theoretical luminosity function is shown in \fig{fakelf}.

To the theoretical luminosity function we fit Gaussians to represent the RC and RGBB:
\begin{align}
\Phirc(\MK) &= \frac{1}{\sigma_{\rm RC} \sqrt{2 \pi}} \exp \left[ -\frac{1}{2} \left( \frac{\MK-\MKRC}{\sigma_{\rm RC}}  \right)^2 \right]  \nonumber  \\
& + \frac{f_{\rm RGBB}}{\sigma_{\rm RGBB} \sqrt{2 \pi}} \exp \left[ -\frac{1}{2} \left( \frac{\MK-\MKRGBB}{\sigma_{\rm RGBB}}  \right)^2 \right]  ~, 
\label{eq:lf}
\end{align}
together with an exponential background of the form of \eqn{bkgd}. While this functional form is formally a poor fit to the Monte Carlo simulated luminosity function, it is motivated by the observed RC magnitude distribution \citep{Alves:00},
and the uncertainties in the isochrones and their metallicity dependence
make a more complex form difficult to justify.
The fit provides us with our fiducial values of the parameters in the luminosity function: $\MKRC=-1.72$, $\sigma_{\rm RC} = 0.18$, $\MKRGBB=-0.91$, $\sigma_{\rm RGBB}=0.19$ and  $f_{\rm RGBB}=0.20$. The K-band magnitude of the RC is slightly brighter than the
value from solar-neighbourhood RC stars ($-1.61$, \citealt{Alves:00,Laney:12}). The width of the RC is in approximate agreement with
the observed distribution \citep{Alves:00}. The RGBB is 0.81 mag fainter
than the RC in the model, and the relative fraction of RGBB stars is
0.2, similar to the I-band fraction \citep{Nataf:13}.
In section \ref{sec:caveats} we investigate the effect of variations from our fiducial form, while at the end of section \ref{sec:3drho} we discuss the effect of spatial variation of the MDF.

In each field we broaden our fiducial luminosity function by the effects of differential extinction and photometric errors.   To do so we measure the standard deviation in the $(\JK)$ colour of each pixel in the extinction map, $\sigma (\JK)$. The spread in $(\JK)$ colour of the red clump results from three factors: 
\begin{inparaenum}[(i)]
\item The intrinsic spread in colour of the red-clump: $\sigma_{\rm RC,JK}$.
\item The photometric error in $J$ and \K, $\sigma_J$ and $\sigma_{\K}$.
\item The residual extinction, $\sigma_{A,JK}$.
\end{inparaenum}

In each pixel we first estimate the residual reddening $\sigma_{E,JK}$ from 
\begin{equation}
\sigma_{E,JK}^2 = \sigma (\JK)^2 - \sigma_{\K}^2 - \sigma_{J}^2 - \sigma_{\rm RC,JK}^2
\end{equation}
where we estimate the intrinsic $\sigma_{\rm RC,JK}=0.05$ from low extinction regions, and $\sigma_{\K}$ and $\sigma_{J}$ are the average photometric errors of the fitted red-clump taken from the VVV DR1 catalog. 

The luminosity function in the \K-band is then broadened from its intrinsic width $\sigma_{\rm RC}(\K)$ by the additional spread due to residual extinction, $\sigma_{A,JK} = A_{\K} \sigma_{E,JK}$, and the photometric error, $\sigma_{\K}$, all added in quadrature. We perform this by convolving our intrinsic luminosity function (equation \ref{eq:lf}) with a gaussian of dispersion $\sigma = \sqrt{\sigma_{\K}^2 +  (\Ak \sigma_{E,JK})^2}$.

\subsection{Deconvolution}
\label{sec:deconv}

The line of sight density distribution is calculated from \eqn{stellarstats} by using a slight variation on Lucy-Richardson deconvolution. Denoting the distance modulus as $\mu \equiv 5 \log ( r/10\pc )$ then \eqn{stellarstats} becomes
\begin{equation}
N(\Kdered) = B(\Kdered) + \Delta \Omega \frac{\ln 10}{5} \int \rhorc (\mu) \Phirc (\Kdered - \mu) r^3 \, d\mu ~.
\label{eq:magdistdistmod}
\end{equation}
Defining in addition $\Delta \equiv \solid (\ln 10/5) \rhorc r^3 $, and denoting convolution as $\star$, then \eqn{stellarstats} is simply
\begin{equation}
N = B + \Delta \star \Phirc ~. \label{eq:Nconv}
\end{equation}
Provided Lucy-Richardson deconvolution \citep{Richardson:72,Lucy:74} converges, it converges to the maximum likelihood deconvolution for Poisson distributed errors, and is therefore an appropriate choice \citep{Shepp:82}. We use a slight variation because while $N(\Kdered)$ is Poisson distributed, 
$N(\Kdered) - B(\Kdered)$ is not.
The equivalent iterative algorithm to Lucy-Richardson in the presence of background can be shown to be
\begin{equation}    
\Delta_{n+1} = \Delta_{n} \left[ \hat{\Phi}_{\rm RC} \star \frac{N}{ (\Delta_{n} \star \Phirc) + B} \right]
\label{eq:lucy}
\end{equation}
where the luminosity function has been normalised, and $\hat{\Phi}_{\rm RC}(M)=\Phirc(-M)$ is its adjoint. Provided $\Delta_n$ converges, then it converges to the maximum likelihood deconvolution of \eqn{Nconv} with Poisson distributed noise in $N$. \Eqn{lucy} reduces to Lucy-Richardson deconvolution in the absence of background (\ie when $B=0$). 

The deconvolution process is carried out on a grid with spacing 0.02\mags. The initial $\Delta$ is a $\sin^2$ (or Hann) function over the range $11.2 < \K < 15$ normalised to give the observed number of background subtracted stars. As is usual with Lucy-Richardson deconvolution, allowing too many iterations results in spurious high frequency structure. We stop the deconvolution when the re-convolved density is consistent with the data. Denoting each of the 0.02\mags bins by a subscript $i$ then our stopping criteria is 
\begin{equation}
\chi^2 \equiv \frac{1}{\nbin} \sum_{i=0}^\nbin \left(\frac{N_i - (\Delta_{n} \star \Phirc)_i - B_i}{\sqrt{N_i}}\right)^2 \leq 1 ~,
\label{eq:stoppingcondition}
\end{equation}
but we impose an upper limit of 100 iterations. In \fig{deconvtest} we show the process of deconvolving one field to obtain the line of sight density. 

\begin{figure*}
\includegraphics[width=150mm]{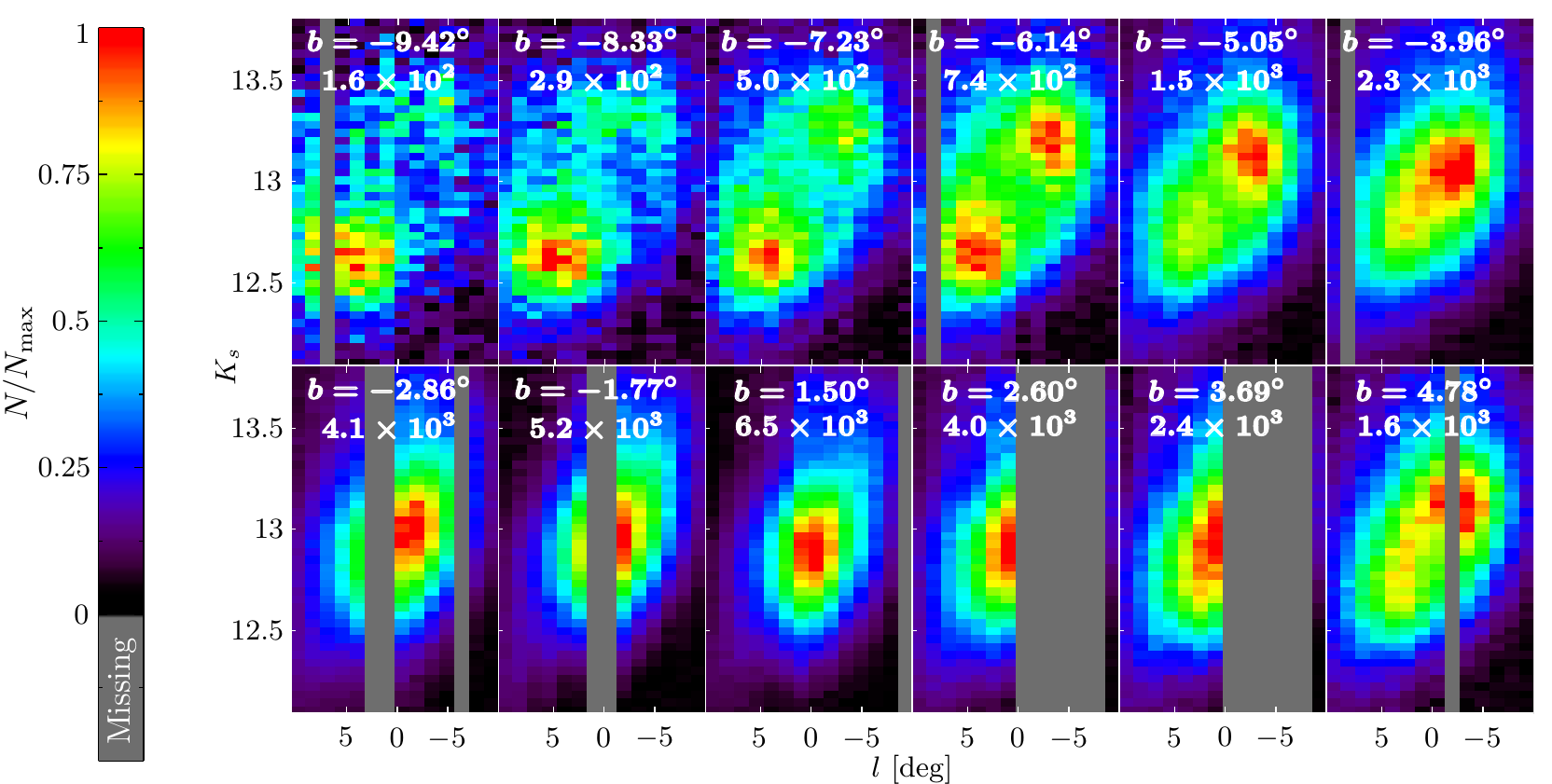}
\caption{Background subtracted histograms for slices at constant latitude. Each VVV field is is halved in latitude giving $\Delta b \approx 0.5\degr$ slices and slices are labeled by their central latitude. Each individual slice is normalised to its maximum, $N_{\max}$, given on each slice below the latitude. For compactness only every other slice is plotted. The histogram bin size is $0.04\mags$.}
\label{fig:histslices}
\end{figure*}

\begin{figure*}
\includegraphics[width=150mm]{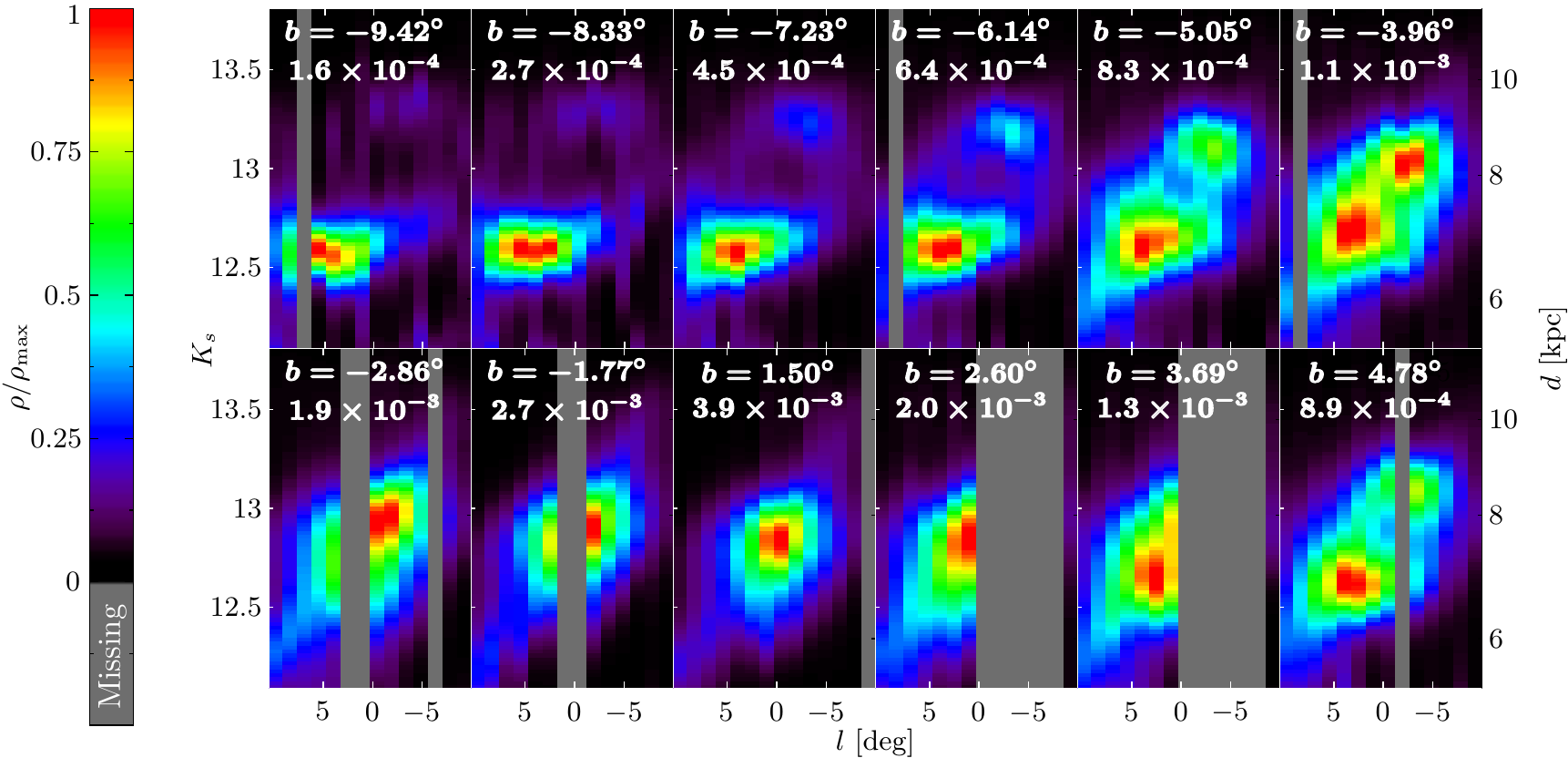}
\caption{Deconvolved density of RC stars, $\rho$. Each VVV field is is halved in latitude giving $\Delta b \approx 0.5\degr$ slices and slices are labeled by their central latitude. The plotted variable is $\rho(r)$ where on the left hand axis $r$ is converted to \K using our assumed $\MKRC=-1.72$. The majority of the difference between this figure and \fig{histslices} is a volume effect: here we plot the density, while in \fig{histslices} we plot histograms. For an ideal standard candle, the number of stars in a histogram binned by magnitude will vary as $\propto \rho r^3$ (\eg~\eqn{magdistdistmod}). The density is therefore lower at relatively larger distances than naive interpretations of \fig{histslices} might suggest. Each slice is normalised to its maximum density, $\rho_{\max}$, given on each slice in ${\rm pc}^{-3}$ below the latitude. For compactness only every other slice is plotted.}
\label{fig:slices}
\end{figure*}

\begin{figure}
\centering
\includegraphics[width=75mm]{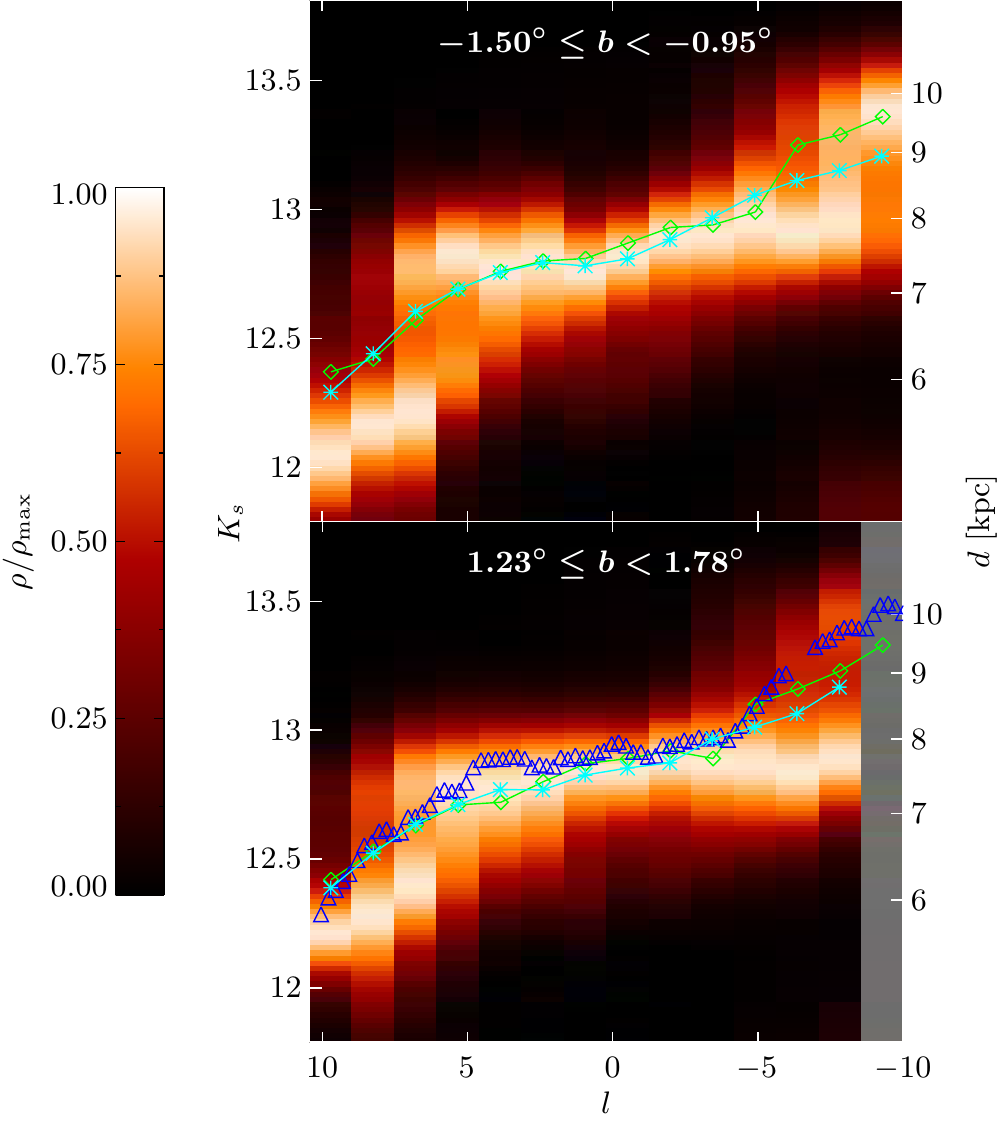}
\caption{Deconvolved line-of-sight density of RC stars compared to the red clump positions found by \citet{gonzalez:11a} (green diamonds) and \citet{Nishiyama:2005hn} (blue triangles). The underlying plotted variable is $\rho(r)$ where on the left hand axis $r$ is converted to \K using our assumed $\MKRC=-1.72$. The line-of-sight density at each $l$ slice is normalised to its peak for easier comparison.  The cyan symbols are the line-of-sight densities converted to the effective measurements of \citet{gonzalez:11a} and \citet{Nishiyama:2005hn}: the $<K_s>$ given by \eqn{meank}.}
\label{fig:oscarcheck}
\end{figure}

In \fig{slices} we show the resultant density obtained by applying the process to all of the fields. The plotted variable is $\rho(r)$ where $r$ is converted to \K using our assumed $\MKRC=-1.72$. Comparing figures \ref{fig:histslices} and \ref{fig:slices} we see that the major result of the deconvolution process is to reduce the far red clump. This is simply the volume effect whereby, for a standard candle, the number of stars in a histogram binned by magnitude will varies as $\propto \rho r^3$.

As a verification we show in \fig{oscarcheck} the computed line-of-sight density for two slices at $l \approx \pm 1\degr$ compared to the measured red clump magnitudes that \cite{gonzalez:11a} and \cite{Nishiyama:2005hn} found at similar latitudes. Again the plotted variable is $\rho(r)$ where $r$ is converted to \K using our assumed $\MKRC=-1.72$. The previously measured red clump locations tend to lie towards the distant edge of the line-of-sight density peak, however this is simply a volume effect \citep{Gerhard:12}: By fitting a Gaussian to the RC luminosity function \cite{gonzalez:11a} and \cite{Nishiyama:2005hn} approximately measure the average red clump \K-band magnitude, weighted by the number of RCGs along each sightline, $\rho r^2$ \citep{Cao:13},
\begin{equation}
<\K> = \frac{\int \K \rho r^2 \, dr}{\int \rho r^2 \, dr} ~.
\label{eq:meank}
\end{equation}
Comparing this to the fitted Gaussian positions of \cite{gonzalez:11a} generally gives very close agreement, as seen in \fig{oscarcheck}.

\section{Three Dimensional Density}
\label{sec:3drho}

\begin{figure}
\centering
\includegraphics[width=50mm]{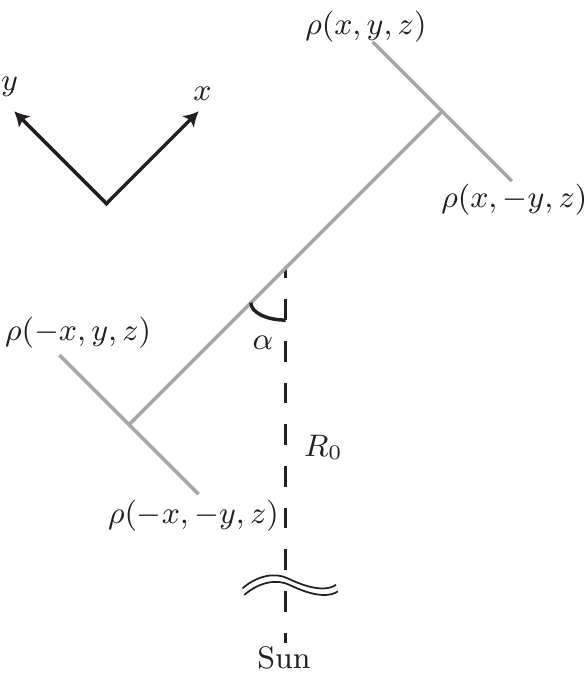}
\caption{Assumed symmetry given a Galactic center distance, \R, and a bar angle, $\alpha$. In addition we assume symmetry about the Galactic plane for a total of 8-fold mirror symmetry. This view is from the North Galactic Pole and the coordinate system is right handed so that $z$ increases out of the page.}
\label{fig:4foldsym}
\end{figure}

In constructing our fiducial three-dimensional density model we assume that the bulge is 8-fold mirror symmetric as illustrated in \fig{4foldsym}. Our reasons for doing so are three-fold: 
\begin{inparaenum}[(i)]
\item VVV DR1 is not yet complete over its survey region, and so the raw density contains `holes' where fields have yet to be observed. Assuming symmetry allows a full three-dimensional model to be constructed.
\item The method results in the averaging of points at different magnitudes, reducing the dependence on the luminosity function and fitted form of the background.
\item The variation in density between the points which are assumed to have equal density allows an error estimate to be easily constructed from their standard deviation. 
\end{inparaenum}

Each deconvolved sightline gives estimates of the density along the line-of-sight, $\rho(l,b,r)$.
The symmetrised density, $\bar{\rho}$, is calculated by first estimating $\rho$ on a 3-dimensional right-handed cartesian grid aligned with assumed axes of symmetry. The coordinates are chosen so that the $x$-axis lies along the major axis of the bar, and the $z$-axis points towards the north Galactic pole (we defer the estimation of these axes until later in this section). The coordinate system is illustrated in \fig{4foldsym}. The cartesian grid spacing, $\Delta x\times \Delta  y\times \Delta  z=0.15 \times 0.1 \times 0.075 \kpc$, was chosen to approximately match the spacing of the measurements in $(l,b,r)$ at 8\kpc. 
The measured densities are then interpolated to the exact grid points using the surrounding points by linearly interpolating the log of the density. Sight-lines not observed in VVV DR1 are considered as missing data and we interpolate only between observed sight-lines. Because our sight-lines exclude $|b| < 1\degr$ we do not have data below $\sim 150\pc$. We therefore do not estimate the density for the lowest $|z|$ slices at $37.5\pc$ or $112.5\pc$ The lowest $z$ slices for which we have calculated the density lie at $|z|=187.50\pc$.

The symmetrised density, $\bar{\rho}$ is then calculated by averaging the measurements from each octant:
\begin{equation}
\bar{\rho}(x,y,z) = \frac{1}{N} \left[ \rho(x,y,z) + \rho(-x,y,z) + \mbox{6 other octants} \right] ~.
\end{equation}
Only grid points with a density measurement are considered observed, and therefore $N$ is less than 8 for many cells.

In order to estimate the axes of symmetry, (\ie the \R and $\alpha$ giving the center and angle to the line-of-sight of the bar) we consider the root mean square deviation from $\bar{\rho}$:
\begin{equation}
\rho_{\rm rms}^2 =  \frac{1}{N} \sum_{\mbox{\small grid points}} \left[ \rho(x,y,z) - \bar{\rho}(x,y,z) \right]^2  ~,
\label{eq:delta}
\end{equation}
where the sum is over the $N$ measured points in the cartesian $x,y,z$ grid. 
We then minimise the quantity
\begin{equation}
\sum_{z=0.4\kpc}^{1\kpc}  \frac{\left<\rho_{{\rm rms}}\right>_z}{\left<\rho\right>_z}
\end{equation}
where $\left<\rho\right>_z$ and $\left<\rho_{{\rm rms}}\right>_z$ are the mean density and mean RMS variation from 8-fold symmetry for a slice at height above the galactic plane, $z$.

We exclude slices with $|z| < 400\pc$.  This is because the slices close to the galactic plane tend to be geometrically thin, while effects such as differential reddening and incomplete or incorrect deconvolution tend to stretch the density peak along the line-of-sight. This line-of-sight stretch corresponds to an artificial reduction in the bar angle to the line of sight, $\alpha$. 

\begin{figure}
\centering
\includegraphics[width=84mm]{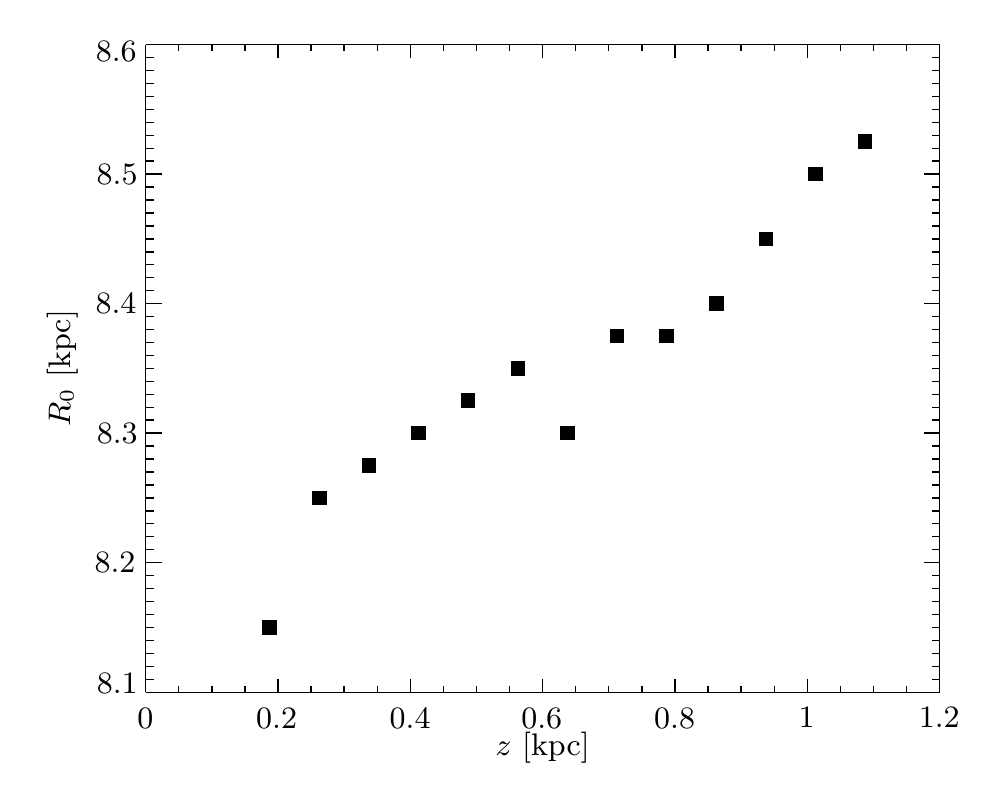}
\caption{The ostensible distance to the galactic center, $R_0$, as a function of the distance from the Galactic plane, $z$ for a bar angle of $\alpha=26.5\degr$. $R_0$ is measured by the minimum departure from 8-fold symmetry, $\rho_{{\rm rms},z}/\left< \rho_z \right>$, for each $z$-slice. A likely explanation is the gradient is the metallicity gradient of the bulge, combined with a slight change in the absolute magnitude of the red clump with metallicity. This is discussed further in the text of \sect{3drho}. }
\label{fig:r0vec}
\end{figure}

We hold $\alpha$ fixed across all $z$ slices, while \R is optimised slice-by-slice. 
We make this choice because at fixed $\alpha$, the distance to the Galactic centre, \R, given by the minimum $\left< \rho_{\rm rms}\right>_z/\left< \rho \right>_z$ varies as a function of $z$. We show this variation at a bar angle $\alpha=26.5\degr$ in \fig{r0vec}. 

The variation in \R of $\approx 0.4\kpc$ corresponds to a change in distance modulus of $\approx 0.1\mags$. The metallicity gradient found by \citet{Gonzalez:13} of $0.28 {\rm dex}/\kpc$ together with the estimated $d\MKRC/d([{\rm Fe}/{\rm H}]) = 0.275$ found by \citet{Salaris:02} would predict a change in red clump magnitude of $0.09$. If correct the apparent change in distance can therefore be explained by a change in the magnitude of the red clump with metallicity. This explanation is degenerate with a steeper extinction law than \citet{Nishiyama:06} at low latitudes, which then flattens at higher latitudes. Although variation of \MKRC due to the MDF gradient seems the more natural explanation than a carefully tailored extinction law, we note that despite several attempts little observational evidence for a non-zero $d\MKRC/d([{\rm Fe}/{\rm H}])$ has been found \citep[\eg][]{Pietrzyski:03,Laney:12}.

Rather than attempting to correct for possible variation of $\MKRC$ we adopt our approach of fixing $\alpha$ across all slices, while optimising \R slice-by-slice, and then simply place each slice in the bar centred coordinates using its \R. This is approximately equivalent to fixing \R and optimising \MKRC.

\begin{figure}
\centering
\includegraphics[width=84mm]{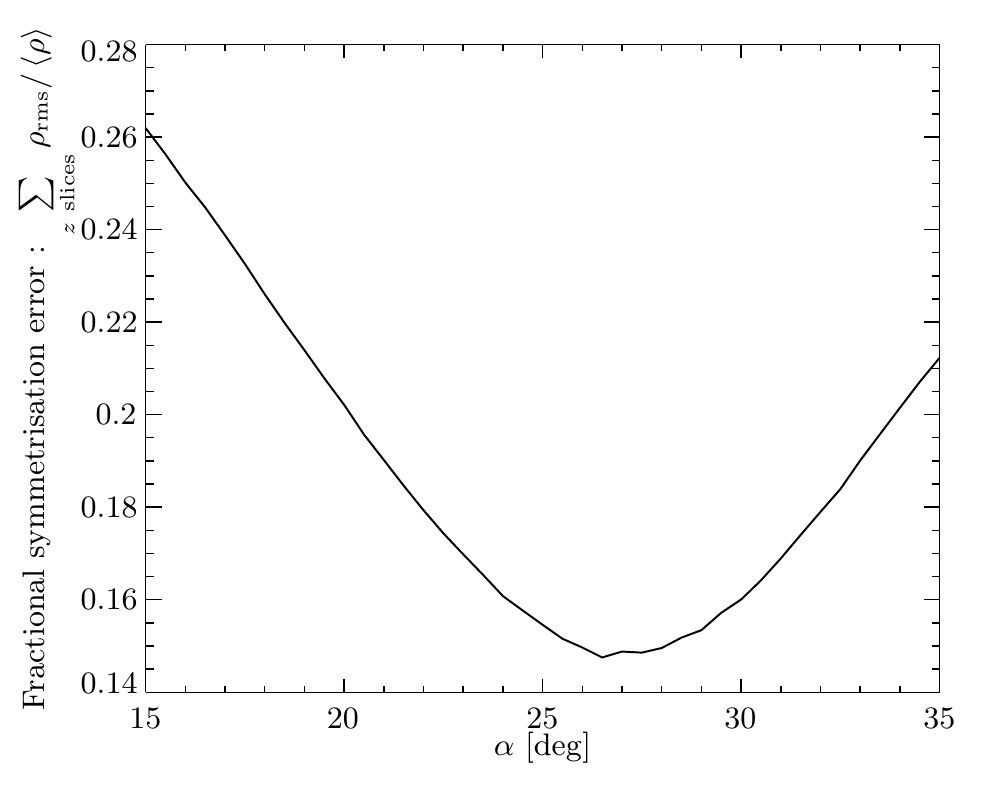}
\caption{The degree of departure from 8-fold mirror symmetry as a function of bar angle to the line-of-sight, $\alpha$. Our measured bar angle is $\alpha = 26.5\degr$, the minimum of this curve.}
\label{fig:findalpha}
\end{figure}

We plot $\sum \left< \rho_{\rm rms}\right>_z/\left< \rho \right>_z$ as a function of the assumed axes of symmetry in \fig{findalpha}. The minimum occurs at $\alpha = 26.5\degr$, and this is our fiducial bar angle.

\begin{figure*}
\includegraphics[width=175mm]{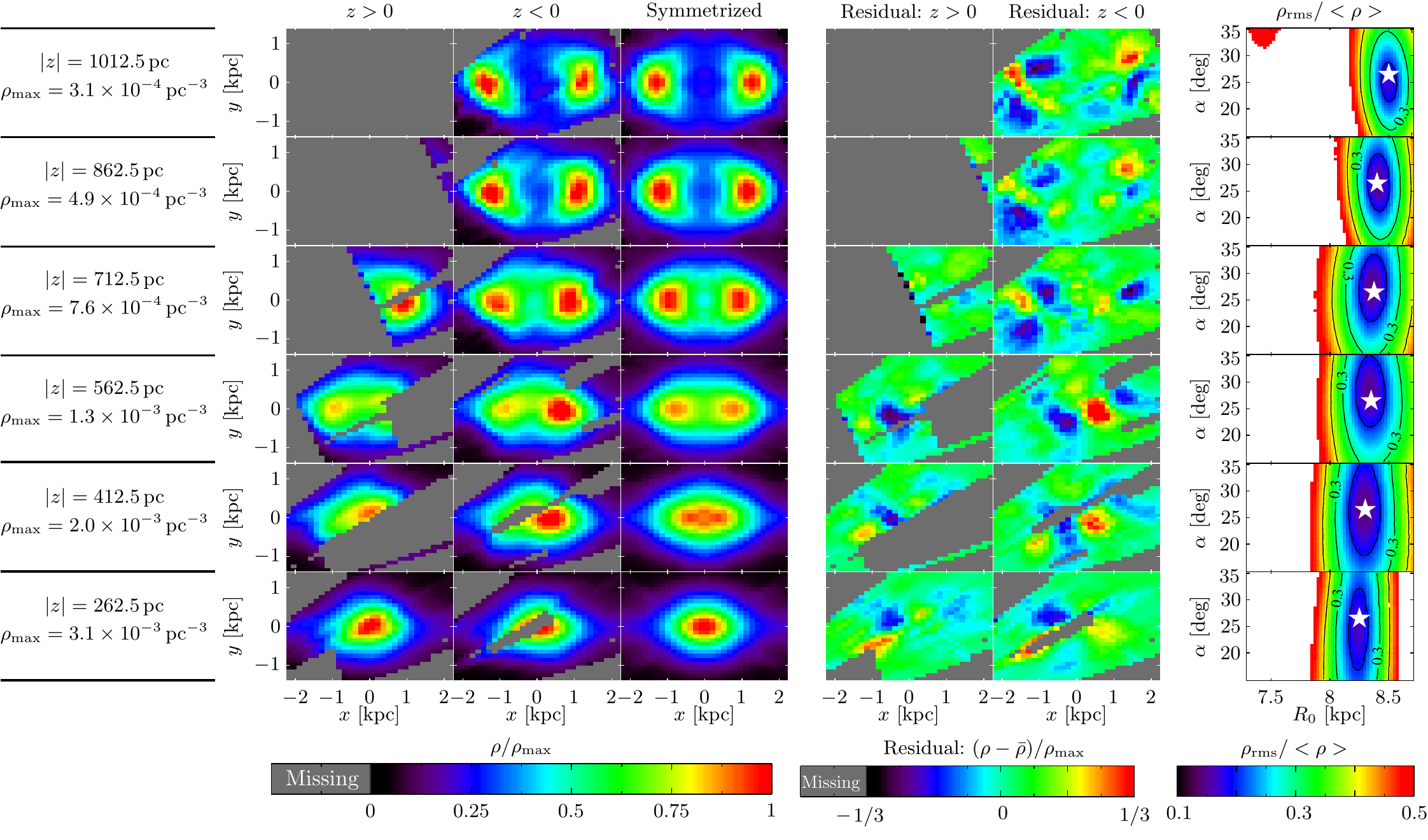}
\caption{Construction of our fiducial three dimensional density measurement. Each row is a different height from the galactic plane, labeled on the left hand side. For compactness we plot only every other slice in the full measurement. The first three columns demonstrate the symmetrisation process. Each row is normalised to the specified $\rho_{\rm max}$. The first two columns are the unsymmetrised data above and below the plane, the third is the symmetrised version of this data. Each point in the third column is therefore the mean of up to eight corresponding points in the first two columns as described in \sect{3drho}. The fourth and fifth columns show the residuals from this symmetrisation process. The final column demonstrates the calculation of the axis of symmetry of the bar used for columns 1-5. The departure from eight fold symmetry, $\left< \rho_{\rm rms} \right>_z / \left<\rho\right>_z$, is plotted as a function of \R and $\alpha$. As described in \sect{3drho}, we hold the bar angle, $\alpha$, fixed across all slices ($26.5\degr$ in this case), but allow \R to vary. In each slice this solution is plotted as a white star. The estimation of $\alpha$ and the variation of \R are shown in greater detail in figures \ref{fig:r0vec} and \ref{fig:findalpha} respectively.}
\label{fig:rhoxyz}
\end{figure*}

We are now able to present the main result of this paper: the fiducial three-dimensional density measurement. This is shown in the third column of \fig{rhoxyz}. In this plot each row is a different $|z|$, the top row being furthest from the galactic plane and bottom row closest to the galactic plane. The columns demonstrate construction of the fiducial density measurement. In the first two columns we plot the raw, unsymmetrised density on bar centred coordinates. The fiducial symmetrised density is then calculated by averaging over the up to 8 available assumed equal points. The fourth and fifth columns show the fractional residuals from this map. The final column shows the symmetrisation error for that slice: ${\left< \rho_{\rm rms} \right>_z }/{\left <\rho \right>_z}$, as a function of $\alpha$ and \R. In this column the white star marks the assumed symmetry whereby $\alpha$ is held fixed across all slices and \R is optimised slice-by-slice. The optimal $\alpha$ and resultant variation in \R for this measurement were those shown in figures \ref{fig:findalpha} and \ref{fig:r0vec}.

\Fig{rhoxyz} shows that the assumption of 8-fold symmetry that underlines the symmetrisation process is generally valid. However a few points are worth noting. In the slice closest to the Galactic plane the bulge appears stretched along the line of sight, despite efforts to reduce this by correcting for effects such as differential extinction. This can be seen in the residual map as an excess of counts stretching towards the sun, and also results in a lowering of the apparent $\alpha$ for this slice. Additionally in several $z<0$ slices there is an apparent asymmetry whereby the far density maximum (large $x$) is $\sim 30$ per cent over populated. This possible asymmetry warrants further investigation, however we defer this until the release of VVV DR2 which will provide data for this region at $z>0$.


\section{Synthetic bulge}
\label{sec:dwek}

In order to verify the reconstruction method and code, it was tested on a synthetic bulge. Magnitude distributions were simulated from a synthetic bulge and disk, which were then used to reconstruct the density of the synthetic bulge.

For the disk density model 2 of \citet{LopezCorredoira:05} was used:
\begin{align}
\rho = \rho_\odot & \exp \left[ - \frac{R-\R}{1970 \pc} - 3740\pc \left(\frac{1}{R} - \frac{1}{\R} \right) \right]  \exp \left[ \frac{|z|}{h_z(R)} \right] ~,   \nonumber 
\end{align}
where $R$ is the Galactocentric distance, and the disk scale height $h_z$ varies as 
\begin{align}
h_z(R) = 285 & \left[  1 + 0.21 \kpc^{-1} (R-\R) +  0.056 \kpc^{-2} (R-\R)^2 \right] \nonumber ~.
\end{align}
The same normalisation, $\rho_\odot=0.05\,{\rm stars}\,\pc^{-3}$, and $K$-band disk luminosity function \citep{Eaton:1984ub} as \citet{LopezCorredoira:05} was also used. 

For the bulge we used one of the analytic models given by \citet{Dwek:95} with parameters recently fit by \citet{Cao:13}. Specifically
\begin{equation}
\rho_{E2} = \rho_{0,{\rm RC}} \exp( -r ) ~, \label{eq:dwekrho}
\end{equation}
where 
\begin{equation}
r = \left[ \left( \frac{x}{x_0} \right)^2 +  \left( \frac{y}{y_0} \right)^2 + \left( \frac{z}{z_0} \right)^2 \right]^{1/2}~,
\end{equation}
and $x_0 = 0.68 \kpc$, $y_0 = 0.28 \kpc$, $z_0 = 0.26 \kpc$ was used. The bar was placed at an angle to the line-of-sight of $\alpha=26.5\degr$ and the normalisation was chosen to be $\rho_{0,{\rm RC}}=0.01\,\mbox{RC stars}\,\pc^{-3}$. This choice of normalisation results in red clump densities similar to those observed in the data.

For the bulge luminosity function \eqn{lf}, our theoretical LF, was used, together with the background fitted from the same Monte Carlo LF simulation.

We then used the density and luminosity functions together with the equation of stellar statistics, \eqn{stellarstats}, to simulate magnitude distributions for each observed sightline. Simulating only observed sight-lines provides a check that the unobserved regions do not introduce spurious features to the density reconstruction. 

\begin{figure}
\centering
\includegraphics[width=84mm]{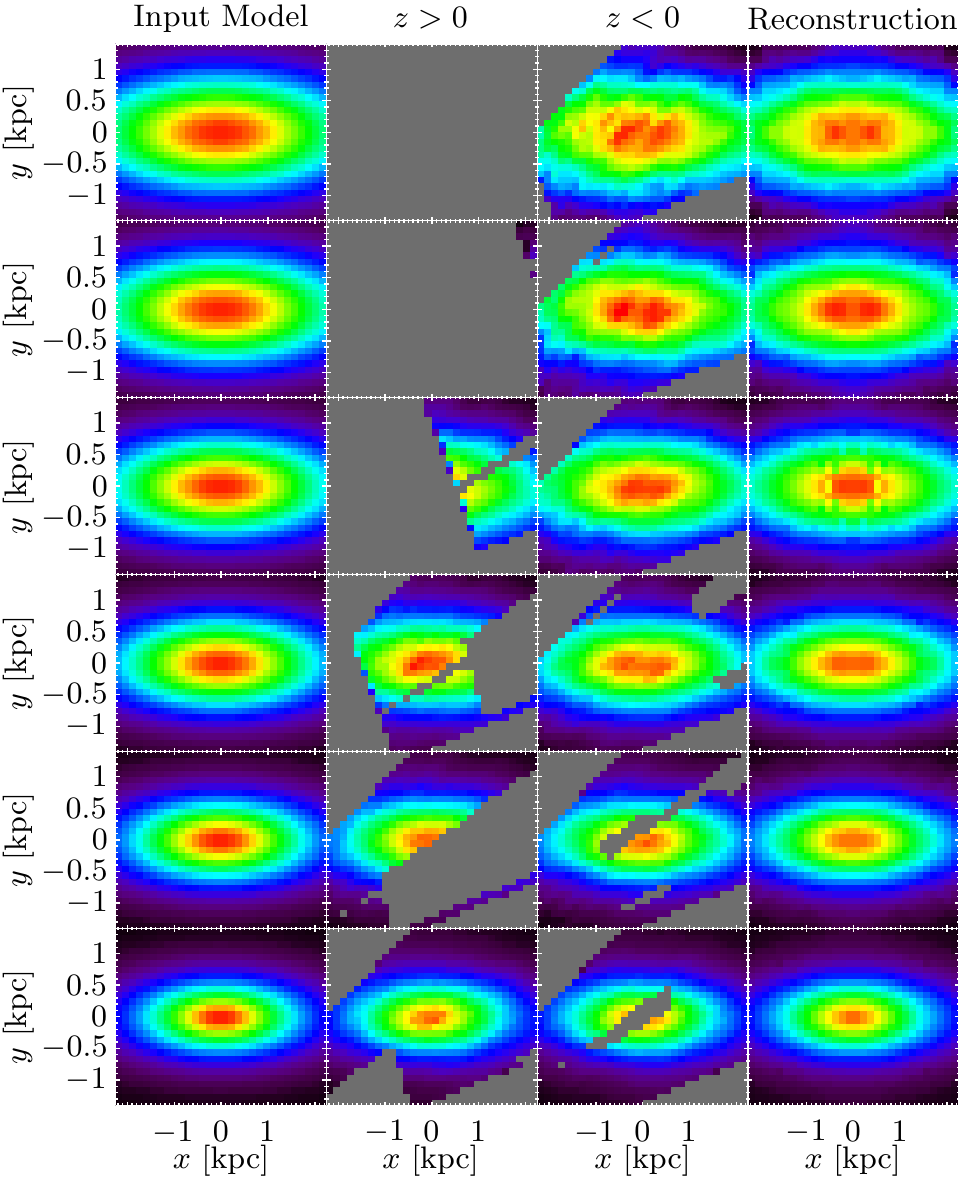}
\caption{Fidelity of the reconstructed synthetic bulge density. The left column shows the input synthetic bulge density. This density is used to produce simulated magnitude distributions in the VVV DR1 sightlines as described in \sect{dwek}. The central two columns show the density reconstructed from these simulations. The right column shows the symmetrised reconstructed density. The same six $z$ slices are shown as in \fig{rhoxyz}. The reconstructed density agrees well with the input model density, and is very different from the density estimated from the data in \fig{rhoxyz}.}
\label{fig:dwekrecon}
\end{figure}

We processed the synthetic magnitude distributions with the same code as was used to process the observed magnitude distributions and construct measured galactic bulge density presented in \fig{rhoxyz}. In \fig{dwekrecon} we show the reconstructed synthetic bulge density side-by-side with the original synthetic bulge density (\eqn{dwekrho}). The recovered bar angle matches the input bar angle exactly, and the mean absolute deviation is less than 10 per cent for all slices with $0.4\kpc < z < 1\kpc$, confirming the accuracy of the reconstruction method.

\section{Systematic variations from fiducial map}
\label{sec:caveats}

In the course of deriving the three dimensional density measurement a number of assumptions were made, particularly:
\begin{inparaenum}[(i)]
\item That the luminosity function is well described by \eqn{lf} with our fiducial parameters.
\item That the background is well represented by \eqn{bkgd}.
\item That the extinction is given by the \citet{Nishiyama:06} extinction law. 
\end{inparaenum}

\begin{figure*}
\includegraphics[width=125mm]{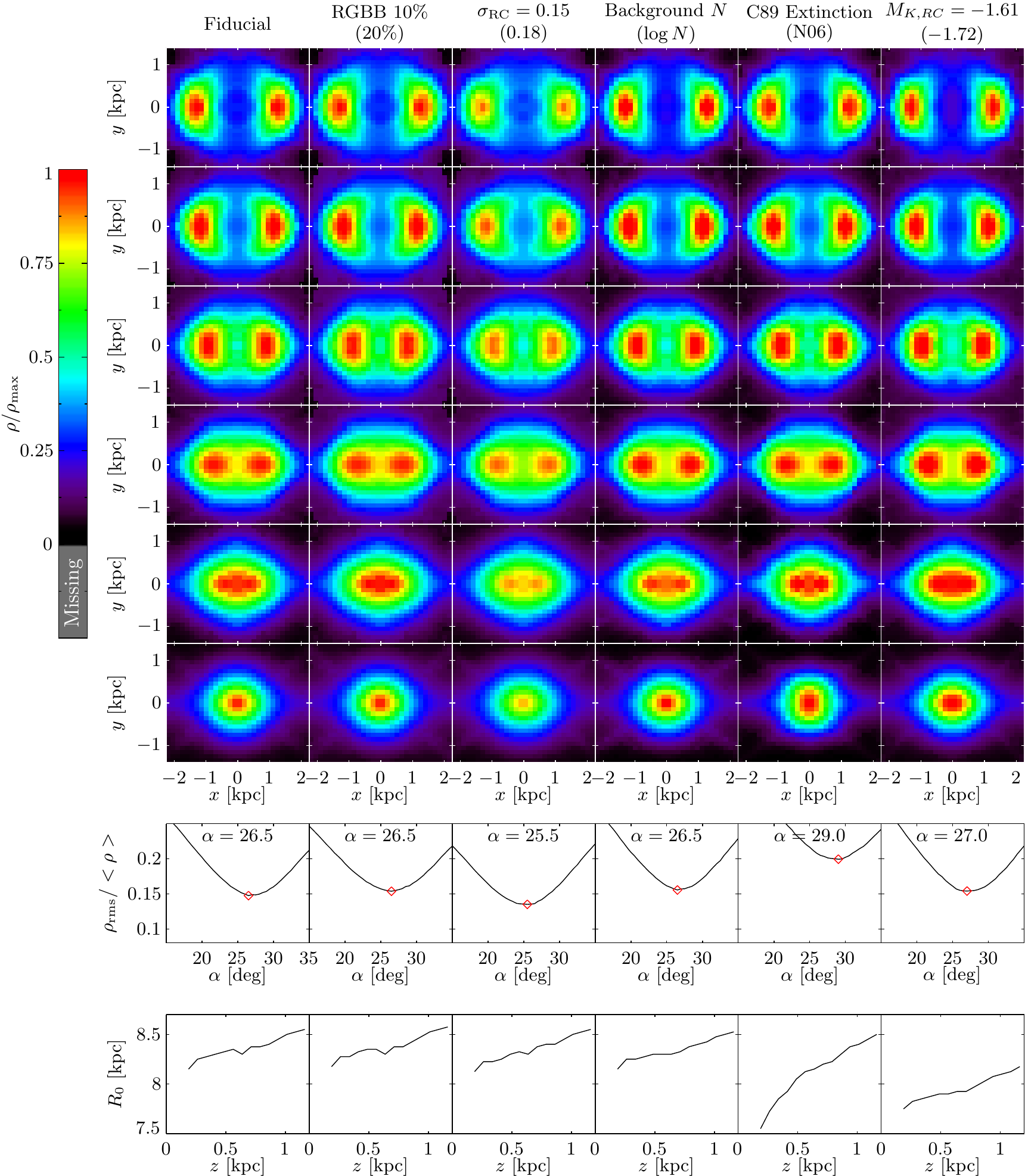}
\caption{The effect of varying the assumptions used to derive the density as described in \sect{caveats}. Each column shows the effect of varying a different assumption. The first column is the fiducial density, columns 2-5 are the variations described as A-E in \sect{caveats}. The upper six rows are the measured symmetrised density constructed as described in \sect{3drho}. The $z$-slices shown are the same as in \fig{rhoxyz}. The lower two rows show calculation of the symmetry axes. The penultimate row shows estimation of the bar angle $\alpha$ similarly to \fig{findalpha}. The estimated bar angle is the minimum of this curve. The final row is the equivalent of \fig{r0vec} and shows the variation of the estimated distance to the galactic center \R as a function of distance from the Galactic plane.}
\label{fig:caveat}
\end{figure*}

Here we test the dependance of the measured density on these assumptions. We perform exactly the same process of measuring the density, under different but reasonable variations in these assumptions. The results of making these variations are shown in \fig{caveat}. Each column shows the results of changing a different assumption. The $z$ slices shown are the same as those in \fig{rhoxyz}. Below this we show the calculation of the axis of symmetry of the bar for each assumption, the equivalent of figures \ref{fig:r0vec} and \ref{fig:findalpha}.
 
In particular, describing each column of \fig{caveat} in turn:
\begin{enumerate}[(A)]
\item We reduce the size of the RGBB relative to the red clump, $f_{\rm RGBB}$, to 0.1 from the fiducial $0.2$. This was motivated by the smaller value of $(0.13\pm0.02)$ originally found using $I$-band data by \citet{Nataf:11}, although this was revised to 0.2 by \citet{Nataf:13b}. The second column of \fig{caveat} shows that the effect of this alteration is minimal and we are therefore robust to this assumption.

\item We reduce the dispersion of the red clump, $\sigma_{\rm RC}$, to 0.15 from the fiducial 0.18. This is motivated by examinations of the red clump in the bulge clusters observed by \citet{Valenti:07,Valenti:10}. Fitting a gaussian together with background to the best defined red clumps gives $\sigma_{\rm RC}$ in the range 0.12 to 0.18. The mean of $\sigma_{\rm RC}$ for these clusters is $0.15$. We consider this a likely lower limit to the width in magnitude of the red clump in the \Kband since the stars in these globular clusters are closer to a simple stellar population than the bulge. Reducing $\sigma_{\rm RC}$ reduces slightly the contrast of the X-shape. This is because, for a single sightline a reduction in $\sigma_{\rm RC}$ will be matched by an increase in the standard deviation of the line-of-sight density. Conserving the number of stars therefore results in a reduction in the density at the peak. In addition the bar angle, 
$\alpha$ is decreased slightly to $25.5\degr$ from $26.5\degr$. This is because increasing the line-of-sight geometric dispersion has given the illusion that the bar is closer to end on.

\item We change the fitted form of the background from a quadratic in $\log N$ (\eqn{bkgd}) to a quadratic in $N$ as often used in RC studies \citep[\eg][]{gonzalez:11a}. While this has a small effect in the high density regions of each slice, there are larger uncertainties towards the edge of each $z$ slice. This is because in these regions the number of red clump stars are smallest compared to the background. We therefore urge caution in interpreting the corners of the map \ie near $|x|=2\kpc$ , $|y|=1\kpc$.

\item We change the extinction law to that given by \citet{Cardelli:89} from our standard extinction law taken from \citet{Nishiyama:06}. 
The major effect of the change in extinction law is in the estimated \R as a function of height. In particular  at small $z$, and with high extinctions, \R is greatly reduced. This is because the \citet{Cardelli:89} extinction law is less steep in the NIR and therefore appears to overcorrect the extinction. It is reassuring therefore that, even using a clearly inappropriate extinction law, the measured density map does not significantly change.

\item We change the magnitude of the red clump to the value measured in the solar neighbourhood by \citet{Alves:00} and \citet{Laney:12} of $\MKRC=-1.61$
 from our fiducial $-1.72$. As expected, the largest change is that \R is shifted by a distance modulus of 0.11, which corresponds 5.2 per cent or $420\pc$ at $8\kpc$. 
We note that measurement of the orbits of the S stars about Sgr A* imply $\R=(8.33\pm 0.35)\kpc$~\citep{Gillessen:09}. Therefore there is presently a small amount of tension with the distance of $\approx 7.9\kpc$ measured using the solar neighbourhood value of $\MKRC=-1.61$ which would be relieved by a slightly brighter \MKRC in the Bulge. 
\end{enumerate}

\begin{figure}
\includegraphics[width=84mm]{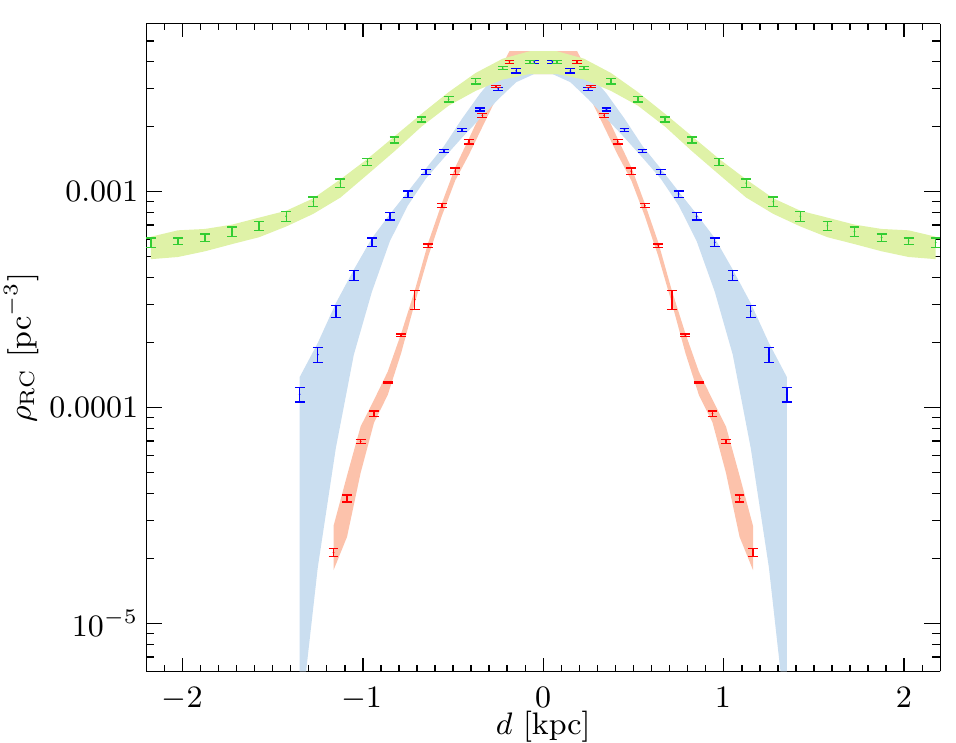}
\caption{Density of red clump stars along the major axis (green), minor axis (red), and intermediate axis (blue). The major axis and intermediate axis are offset from the Galactic plane by 187.5\pc. The error bars are the internal errors: the standard error on the mean of the symmetrisation process described in \sect{3drho}. The shaded regions are estimated systematic errors corresponding to the range of densities calculated in \sect{caveats}. The measured density is typically accurate to $\sim 10$ per cent, the exception is at large values of $y$ along the intermediate axis where uncertainties in the background fitting and extinction dominate (variations C and D in \sect{caveats}).}
\label{fig:axis}
\end{figure}

In \fig{axis} we plot the density along the major, intermediate and minor axis of the bar \ie along the $x$, $y$ and $z$ axes. The major and intermediate axis are offset from the plane by $187.5\pc$ since we do not have density measurements in the plane. The points are the measurements in the fiducial model and their error bars are the standard error on the mean calculated from the symmetrisation process. We consider these to be the internal errors of the density measurement process. The shaded region is the range of densities spanned by the densities recovered using assumptions A-E. We consider these to be an estimate of the systematic errors arising from the deconvolution process.

We conclude from figures \ref{fig:caveat} and \ref{fig:axis} that, despite the assumptions needed in the measurement process, the density tends to be robust at the $\sim \pm 10$ per cent level. The exception being furthest along the intermediate axis of the map where the level of background is high compared to red clump stars. In these regions the fitted background form in particular has a larger effect.

From these experiments we also conclude that $\alpha=(27\pm2)\degr$, where the error spans the range given by varying assumptions A-E. This is consistent with the recent parametric measurements using red clump stars of $\alpha=24-27\degr$ by \citet{Rattenbury:07} and marginally consistent with $\alpha=29-32\degr$ found by \citet{Cao:13}.



\section{Projections}
\label{sec:proj}
\label{sec:axis}

\begin{figure}
\includegraphics[width=84mm]{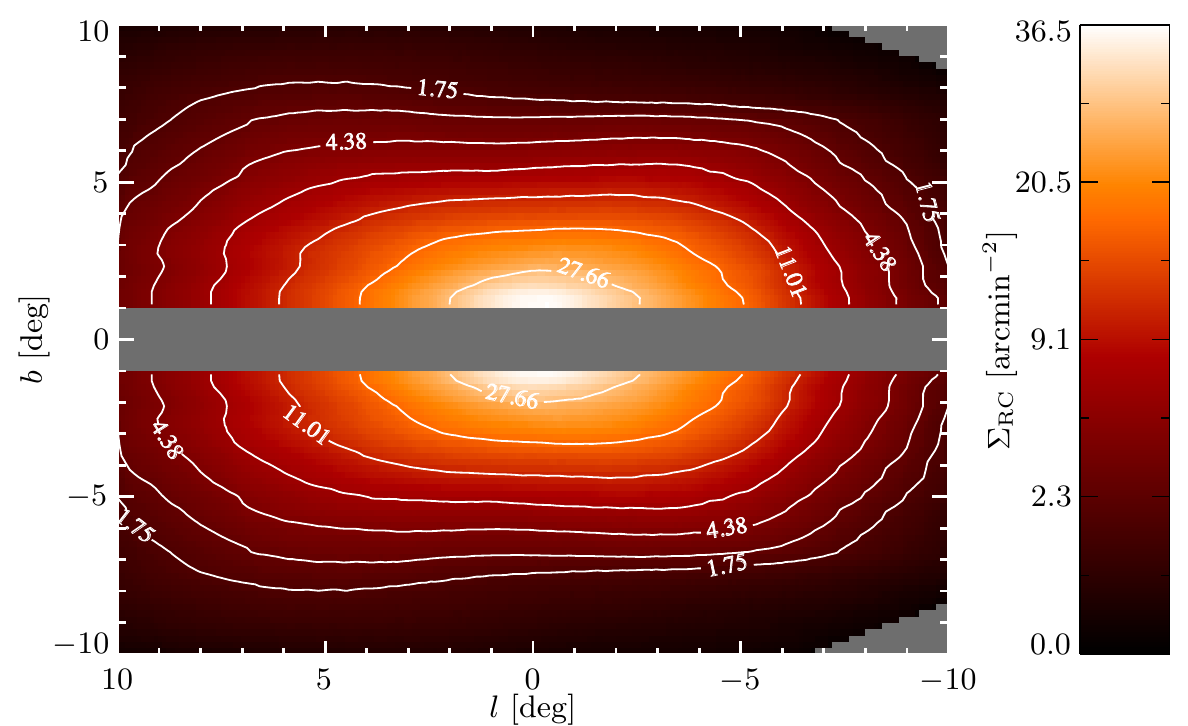}
\caption{The three dimensional density model re-projected to the surface number density of red clump stars visible from the sun. Contours represent isophotes separated by 0.5\mags. In the re-projection of the model the bulge was assumed to lie at 8\kpc. The log density was linearly interpolated to a grid in $l$ and $b$ before integrating $\rho r^2$ along each simulated $l,b$ sightline.}
\label{fig:sigmaproj}
\end{figure}

\begin{figure}
\includegraphics[width=84mm]{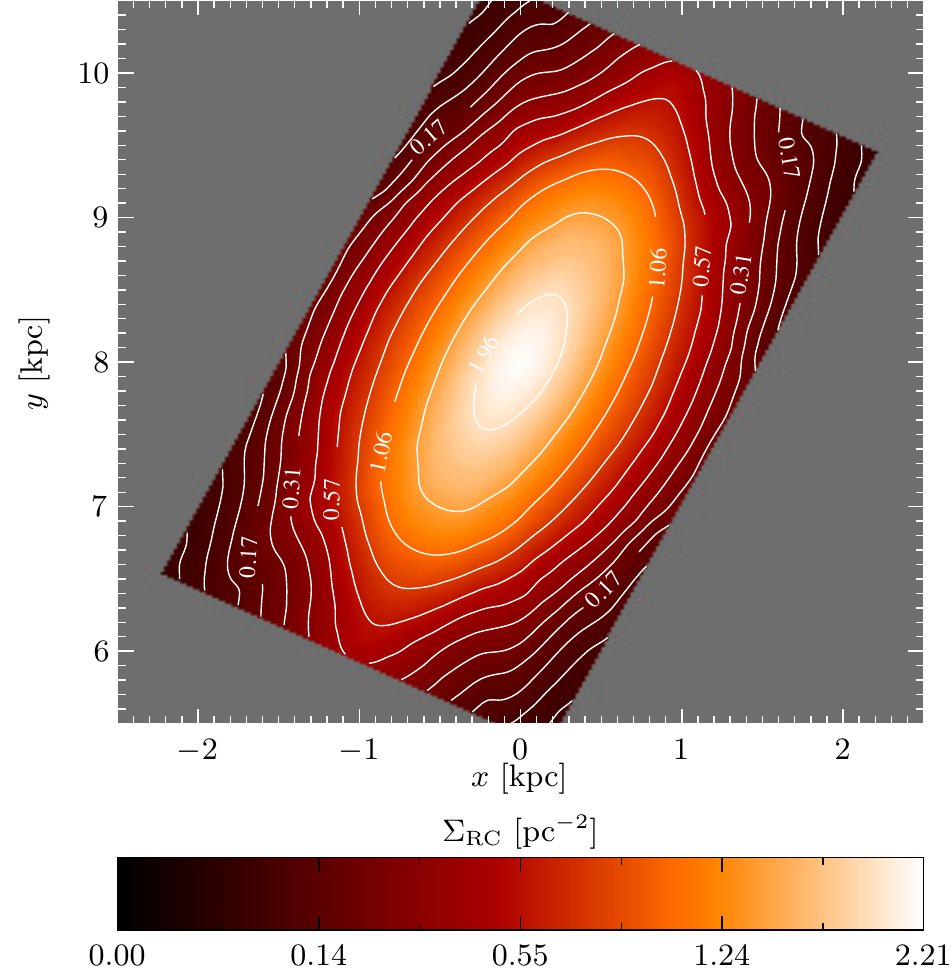}
\caption{The Milky Way Bulge viewed from above \ie the three dimensional density measured in this work projected from the north Galactic pole. Numbers give the surface density of red clump stars in ${\rm pc}^{-2}$, contours define isophotes separated by ${}^1/_3$\mags. The extinction within $150{\rm pc}$ of the galactic plane is too high for reliable density measurements, and is therefore excluded from the projection. The maximum extent above the plane is $1150{\rm pc}$.}
\label{fig:aboveproj}
\end{figure}

\begin{figure}
\includegraphics[width=84mm]{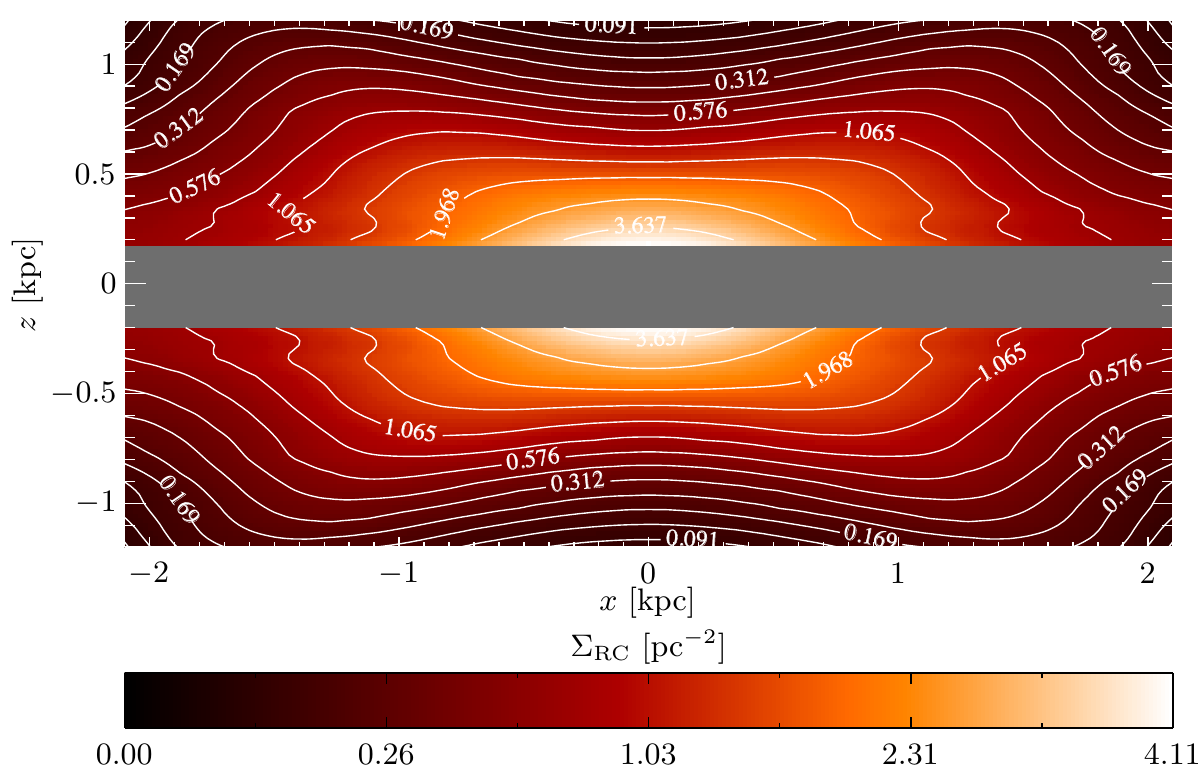}
\caption{The three dimensional density of the Milky Way Bulge measured in this work projected along the intermediate axis. Numbers give the surface density of red clump stars in ${\rm pc}^{-2}$, contours define isophotes separated by ${}^1/_3$\mags. The density map extends to 1.4\kpc along the intermediate axis and therefore these are the limits of integration in the projection.}
\label{fig:sideproj}
\end{figure}

In figures \ref{fig:sigmaproj}, \ref{fig:aboveproj} and \ref{fig:sideproj} we show three projections of the fiducial density measurement. 

In \fig{sigmaproj} we show the projection from the sun. We show this projection since the surface density of bulge stars is easily accessible to other observations and methods. We note that in this work we have measured the density of red clump stars. Comparison of this projection to other measurements allows the fidelity of red clump stars to trace the underlying density to be assessed. The projection was constructed by linearly interpolating the log density to a regular $\rho(l,b,r)$ grid and integrating $\rho r^2$ along each simulated $(l,b)$ sightline.

\Fig{aboveproj} shows the projection of the density from the Galactic North Pole. The projection was constructed by integrating over $z$ and therefore excludes $|z| > 1150 {\rm pc}$ and $|z| < 150{\rm pc} $. For visualisation purposes this is then tilted to our measured bar angle of $\alpha = 26.5\degr$ and  up-sampled via fitting the minimum curvature surface (\citealt{Franke:82}, implemented in \textsc{IDL} routine \verb|MIN_CURVE_SURF|) to the log surface density. Measuring the ellipticity of the isodensity curves in this projection gives values between 0.4 and 0.5, becoming slightly more circular towards the center. Care should be taken in interpreting this however since the projection excludes $|z| < 150{\rm pc}$ \citep[\cf figure 3 of][]{Gerhard:12}.

\begin{figure}
\includegraphics[width=84mm]{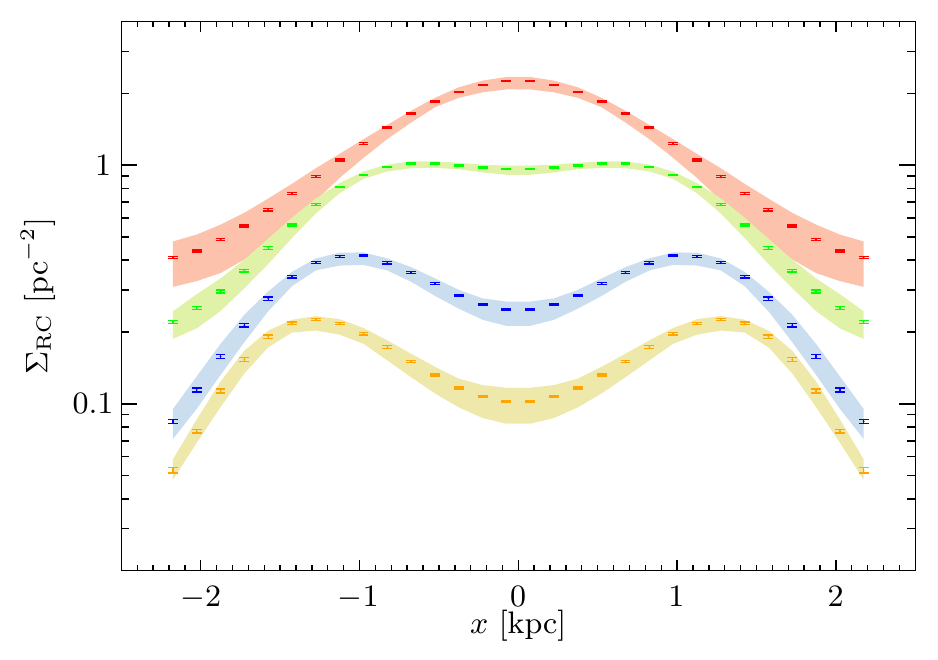}
\caption{Surface density of red clump stars for the side view, \fig{sideproj}, averaged over slices at $0.15\leq z/\kpc<0.45$ in red, $0.45 \leq z/\kpc< 0.75$ in green, $0.75\leq z/\kpc<1.05$ in blue, and $1 \leq z/\kpc< 1.2$ in yellow. As in \fig{axis}, error bars are the internal errors from the symmetrisation process, and the shaded region are the estimated systematic errors estimated from the variations in \rhorc~caused by the changes considered in \sect{caveats}.}
\label{fig:inmaslice}
\end{figure}

\Fig{sideproj} shows a projection of the density along the intermediate axis \ie a side on view of the Milky Way Bulge. It was produced from the measured density in the same manner as \fig{aboveproj} but integrating over the $y$ direction. In \fig{inmaslice} we plot slices at different heights above the plane through \fig{sideproj}. Both the side on view in \fig{sideproj} and the surface density slices in \fig{inmaslice} vividly illustrate the strong X-shape of the galactic bulge for $|z| \gtrsim 0.5\kpc$. In principle these can be compared to images of edge on galaxies and N-body models such as \citet{Bureau:05}, \citet{Bureau:06}, and \citet{Inma:06}. Care must be taken however, since the integration includes only the region along the line-of-sight within $\pm 1.4\kpc$ of the bulge, and therefore excludes any disk component.

\section{Discussion}
\label{sec:conc}

In this work we have presented the first direct non-parametric three-dimensional measurement of the density of red clump stars in the galactic bulge. Most works have generally either fitted parametric models of the bulge \citep[\eg][]{Dwek:95,Rattenbury:07,Cao:13} or attempted to non-parametrically deproject the bulge from surface brightness data \citep[\eg][]{Binney:97,Bissantz:02}. The exception is \citet{LopezCorredoira:05} who inverted star counts brighter than the red clump towards the bulge. However, the red clump provides a better standard candle resulting in a higher resolution and a more direct reconstruction.
 
The resultant three dimensional density of the bulge prominently displays a boxy/peanut-like structure, qualitatively very similar to that predicted by N-body simulations of the buckling instability \citep[\eg][]{Athanassoula:02,Inma:06}. In particular the side view presented in \fig{sideproj} is qualitatively very similar to those presented in \citet{Inma:06}. 

Fitting exponentials to the density profiles of the major, intermediate and minor axes shown in \fig{axis} between 0.4\kpc and 0.8\kpc gives scale lengths of $0.70\kpc$, $0.44\kpc$ and $0.18\kpc$ respectively. This corresponds to axis ratios $10:6.3:2.6$, however a single axis ratio cannot fully describe the data along these axes, nor can it describe the X/peanut shape. The minor axis scale length is shorter than usually found when fitting parametric models \citep{Cao:13,Rattenbury:07}, however this is a result of the X-shape. Performing the same exercise of fitting an exponential at distances $0.525,1.125$ and $1.725\kpc$ along the major axis gives vertical scale heights of $0.25,0.56$ and $0.46\kpc$ respectively.

The vertical symmetry of the X-shape in the unsymmetrised density shown in \fig{rhoxyz} and between positive and negative latitudes in \fig{slices} suggests that the galactic center is not currently undergoing a buckling episode and instead is secularly evolving after buckling \citep{Inma:08,Athanassoula:08}. We defer quantitative investigation of the top/bottom asymmetry to the release of VVV DR2 where the large region of missing fields above the plane will have been observed.
We note that this conclusion is also qualitatively supported by the symmetry about the plane in the COBE-DIRBE data \citep{Weiland:94} as well as by the prominence of the X-shape in the density, as shown in \fig{sideproj}, all suggesting a similar conclusion that the bulge is secularly evolving after buckling.

We hope that the 3D-density measurement of the Bulge presented here will be useful for a variety of purposes, for example: 
\begin{inparaenum}[(i)]
\item Constraining dynamical models of the Milky Way such as those produced in \cite{Shen:10} and \citet{MartinezValpuesta:11}.
\item Investigating the orbital distributions of the different stellar populations found in spectroscopic surveys of the Bulge \citep{Babusiaux:10,Ness:13b}
\item Building population models such as the Besan\c{c}on models \citep{Robin:03}.
\item Examining the gas flow in the potential implied by the density measurement.
\item Computing microlensing optical depths and statistics towards the bulge, important for example in planning and understanding the Euclid and WFIRST microlensing surveys \citep{Beaulieu:10}.
\end{inparaenum}

Ultimately we hope that the results presented will help towards a detailed understanding of the formation and evolution of the Milky Way Bulge, for which we can resolve the constituent stars, and which we can therefore study in greater detail than bulges in external galaxies.

\section{Acknowledgments}

We gratefully acknowledge useful discussions with Oscar Gonzalez, Manuela Zoccali, Matthieu Portail, and especially Inma Martinez-Valpuesta.

We are indebted to the VVV team for providing user friendly images, catalogs, and the information needed to utilise them, both online and in the DR1 survey paper, \citet{Saito:12}.

This work heavily relied on DR1 of the VVV survey provided as Phase 3 Data Products by the ESO Science Archive Facility.
Based on data products from observations made with ESO Telescopes at the La Silla or Paranal Observatories under ESO programme ID 179.B-2002.
{\footnotesize
\expandafter\ifx\csname bibcodeprefix\endcsname\relax\def\bibcodeprefix{}\fi
\def\bibcode#1{\href{http://adsabs.harvard.edu/abs/#1}{\texttt{[ADS]}}}

}
\bsp
\label{lastpage}
\end{document}